\documentclass[fleqn,usenatbib]{mnras}
\usepackage{newtxtext,newtxmath}
\usepackage[T1]{fontenc}
\usepackage{ae,aecompl}

\usepackage{graphicx}	
\usepackage{amsmath}	
\usepackage{amssymb}	
\usepackage{colordvi}

\title[DLAs and HI in galaxies: the GAEA view]{Damped Lyman-$\alpha$ absorbers and atomic hydrogen in galaxies: the view of the GAEA model}

\author[S. Di Gioia et al.]{
Serafina Di Gioia,$^{1,2,3,4}$\thanks{E-mail: serafina.digioia@inaf.it}
Stefano Cristiani,$^{2,3,4}$
Gabriella De Lucia,$^{2}$
Lizhi Xie$^{5}$
\\
$^{1}$ Dipartimento di Fisica dell' Universit\`a di Trieste, Sezione di Astronomia, via Tiepolo 11, I-34131 Trieste, Italy\\
$^{2}$ INAF--Osservatorio Astronomico di Trieste, Via G.B. Tiepolo, 11, I-34143 Trieste, Italy \\
$^{3}$ IFPU--Institute for Fundamental Physics of the Universe, via Beirut 2, I-34151 Trieste, Italy \\
$^{4}$ INFN-National Institute for Nuclear Physics, via Valerio 2, I-34127 Trieste\\
$^{5}$ Tianjin Normal University, Binshuixidao 393, Tianjin, China}

\date{Accepted XXX. Received YYY; in original form ZZZ}

\pubyear{2020}

\begin{document}
\label{firstpage}
\pagerange{\pageref{firstpage}--\pageref{lastpage}}
\maketitle

\begin{abstract}
Using the GAEA semi-analytic model, we analyse the connection between Damped Lyman-$\alpha$ systems (DLAs) and HI in galaxies. Our state-of-the-art semi-analytic model is tuned to reproduce the local galaxy HI mass function, and that also reproduces other important galaxy properties, including the galaxy mass - gas metallicity relation. 
To produce catalogs of simulated DLAs we throw $10^5$ random lines of sight in a composite simulated volume: dark matter haloes with log$(\frac{M_{200}}{ M_{\odot}}) \geq 11.5$ are extracted from the Millennium Simulation, while for  $9.2 \leq \log(\frac{M_{200}}{ M_{\odot}})<11.5$  we use the Millennium II, and for $8 \leq \log(\frac{M_{200}}{M_{\odot}}) < 9.2$ a halo occupation distribution model.
At $2 < z < 3$, where observational data are more accurate, our fiducial model predicts the correct shape of the column density distribution function, but its normalization falls short of the observations, with the discrepancy increasing at higher redshift. The agreement with observations is significantly improved increasing both the HI masses and the disk radii of model galaxies by a factor 2, as implemented 'a posteriori' in our $2M-2R$ model. In the redshift range of interest, haloes with $M_{200} \geq {10}^{11} M_{\odot}$ give the major contribution to $\Omega_{\rm DLA}$, and the typical DLA host halo mass is $\sim {10}^{11} M _{\odot}$. The simulated DLA metallicity distribution is in relatively good agreement with observations, but our model predicts an excess of DLAs at low metallicities. Our results suggest possible improvements for the adopted modelling of the filtering mass and metal ejection in low-mass haloes.

\end{abstract}




\section{Introduction}

In the modern cosmological framework, large-scale structure develops hierarchically, due to the growth of gravitational instabilities in a fluid dominated by dark matter and dark energy \citep[e.g.,][]{Peebles1981JBAA,Peebles1984,Springel2003MNRAS}. Galaxies form and evolve in this cosmological scenario, and it is nowadays accepted that a crucial element to understand the physical processes driving galaxy evolution is cold gas. In particular, the cold gas distribution in galaxies at different cosmic epochs should be quantified, understanding how galaxies accrete and lose their gas as a function of cosmic time and environment. These questions have been subject of intense research activities in the past decades
\citep{Silk-Mamon2012RA,Conselice2013MNRAS,Fraternali2014IAUS,Spring2017MNRAS, Sorini2018ApJ,Whitney2019ApJ}.

With present and upcoming facilities (e.g. MUSE, ALMA, ELT), allowing us to trace the gaseous components of galaxies out to their outskirts, this is an ideal time to study the cycle of gas (and metals) in and around galaxies.
Hydrogen is the most abundant element in the galactic cold phase, 
and can be detected in emission (21cm line -  
mostly in the local Universe)  
or in absorption (Lyman-$\alpha$ line, in optical for $z \geq 1.65$ and in UV for lower redshifts).
Due to the sensitivity of current instrumentation, the detection in emission (21cm line) is strongly biased towards the brightest galaxies/highest column densities, and is limited to relatively low redshift (up to $z=0.06$).
In the last decade, the HI 
content of galaxies has been characterized for a large sample of local galaxies thanks to surveys like HIPASS, ALFALFA, GASS 
\citep[][]{Meyer-HIPASS2004MNRAS,Giovanelli-ALFALFA2005AJ,Catinella2010GALEX1,Catinella2013ARECIBO-groups,Catinella2018a}.
These surveys have also allowed studies of the correlation between HI and galaxy stellar mass or other galaxy properties (e.g. star formation rate, environment, etc.). 

Studies based on absorption lines are not affected by the same observational limits of emission line studies: the Lyman-$\alpha$ line results from a transition between the $2^2$ P state and the $1^2$ S (ground) state of the hydrogen atom ($\lambda = 1215$ \AA), and it is possible to observe it from the ground at z$\gtrsim 1.6$. The systems characterized by the strongest absorption lines are the Damped Lyman-$\alpha$ systems (DLAs), defined as hydrogen absorbers with column density $N _{\rm H I} > {10}^{20.3} $ atoms cm$^{-2}$.
These strong absorbers are typically associated with low-ionization metal line complexes  \citep{Prochaska2003ApJS, Noterdaeme2012,Rafelski2012ApJ}, suggesting that they are part of a gaseous medium affected by chemical enrichment, like the ISM and the CGM in galaxies. 

Large spectroscopic surveys, such as the Sloan Digital Sky Survey (SDSS; \citealt{Schneider2010AJ}) and BOSS \citep{Eisenstein2011AJ-BOSS}, have greatly improved the statistics for samples of high-redshift absorbers ($1.5 <z < 4.5$), tightening the constraints on the shape of the column density distribution function, the comoving line density of
DLAs, and the evolution of the neutral gas density \citep[e.g][]{Storrie-Wolfe2000ApJ,Peroux2003,Noterdaeme2012,Crighton2015}. These studies have demonstrated that DLAs contain $\sim 80 \% $  of the neutral  gas  available  for  star  formation  \citep{Prochaska2009,Noterdaeme2012, Zafar2013A&A,Storrie-Wolfe2000ApJ,Peroux2003,Prochaska_2005}, so DLAs studies provide us with an
estimate of the gas available for star formation from $z=5$ to now.

\cite{Rafelski2012ApJ} and \cite{Neeleman2013ApJ} have estimated the metallicities for a sample of DLAs in the redshift interval $(2<z< 4)$, and investigated their mean metallicity evolution. Recently, these measurements have been updated by \citet{DeCia2018} who developed a procedure to estimate DLA metallicities corrected for dust depletion. 

In the last decades, numerous follow-up observations have been carried out to identify the counter-parts of DLAs, mainly at low redshift \citep[e.g.][]{Chen2003ApJ,Rao2011MNRAS_DLAcounterparts,Rahmani2016MNRAS_DLAcounterparts}. Different techniques have been used: narrow-band imaging of the fields around the background quasar \citep[][]{ MollerWarren1998,Kulkarni2007ApJ,Fumagalli2010MNRAS,Rahmani2016MNRAS_DLAcounterparts}, 
long-slit spectroscopy to search for emission lines from the galaxy associated with the DLA system \citep[e.g.][]{Moller2002ApJ,Fynbo2010MNRAS,Fynbo2011MNRAS,Noterdaeme2012,Srianand2016MNRAS,Krogager2017}, 
integral field spectroscopy 
\citep[][]{Peroux2011MNRAS_SINFONI,Wang-Kanekar2015MNRAS}, and sub-millimeter observations with ALMA \citep{Neeleman2019ApJ}.
The detection rate in blindly selected samples remains very low \citep{Fumagalli2015ObservationsonDLAsCOUNTERPARTS-doubleDLAtechnique}, but increases when strong cuts on the DLA metallicity are applied \citep{Krogager2017}. These results suggest that DLAs are likely associated with low-luminosity galaxies, most of which are below current observational capabilities \citep{Krogager2017}.

The occurrence of strong HI absorbers detected at high impact parameters ($b> 30$kpc) from their likely host galaxies  \citep{Christensen2019arXiv190805363C,MC2019,Peroux-MUSE-CGM} provides insight into their origin and clustering properties.
While in early DLA studies it was commonly believed that they originate from the absorption of gas settled in the disks of massive galaxies \citep[][]{ProchaskaWolfe1997ApJ}, there is now ample observational evidence that small and intermediate mass galaxies provide a non negligible contribution to DLA statistics \citep{Krogager2017}, in accordance with the predictions of the theoretical study by \cite{Rhamati2014}.

\cite{Font-Ribera2012} carried out a cross-correlation analysis of DLAs (selected from the BOSS survey) with the Lyman-$\alpha$  forest and obtained constraints on the DLA cross-section as a function of halo mass. The bias they find implies a typical DLA host halo mass of $\sim {10}^{12} M_{\odot}$ at $z=2$. In 2018 \cite{Perez-Rafols2018MNRAS} updated the results by \cite{Font-Ribera2012} finding a typical DLA halo mass of $\sim 4 \times {10}^{11} M_{\odot}$. In the meantime, \citet{Arinyo-i-prats2018} developed a new method to classify the metal strength of DLAs and studying the dependence of the bias on the metallicity of the absorbers  \cite{PerezRafols18} showed that the linear bias associated with DLAs decreases as their metallicity decreases. 

In the last 20 years, a number of theoretical studies have used hydro-dynamical simulations to investigate the nature of strong HI absorbers and DLAs in particular \citep[e.g.][]{Gardner1997ApJ,Gardner2001,Haenelt1998ApJ,Nagamine2004MNRAS.348..421N,Pontzen2008,Tescari2009MNRAS,Razoumov_2009,Fumagalli2011MNRAS,Cen2012ApJ,VanDeVoort2012DLAs,Altay2013,Rhodin2019NatureofDLAs,Hassan2020MNRAS.492.2835H}. The resolution of the simulations adopted has increased over time, but the approach typically needs to resort to different layers of sub-grid prescriptions to model the high HI column densities of DLAs. 
Some studies overcome the absence of a full cosmological distribution of absorbers by combining results from small-scale simulations with analytic parametrizations of the halo mass function to predict statistical properties of the DLA population \citep[e.g.][]{Gardner1997ApJ,Gardner2001}, or to study the nature of the host galaxies \citep[e.g.][]{Pontzen2008}. This approach can lead to biased results, requiring some strong assumptions about the environments that can give rise to DLA absorbers. In addition, it does not account for the potentially large scatter in the distribution of absorbers for haloes of similar properties. 

Studies based on hydro-dynamical simulations have pointed out an important contribution to the DLA population, typically increasing with increasing redshift, from gas that is not associated with the ISM of galaxies. There is no consensus on the quantitative estimate of such a contribution that ranges, depending on the study, between $\sim 20$ per cent \citep{VNavarro2018} to more than $\sim 50$ per cent \citep{Fumagalli2011MNRAS,VanDeVoort2012DLAs}. 
Most numerical studies indicate a major contribution to the DLA population at $2 \leq z \leq 3$ from haloes with virial masses of $10^{10} - {10}^{12} M_{\odot}$ \citep{Cooke2006ApJ,Pontzen2008,Barnes2009MNRASB,Font-Ribera2012}.

In this study, we focus on an alternative theoretical approach provided by semi-analytic models of galaxy formation. While these are unable to resolve the internal structure of galaxies and do not model the hydro-dynamical processes self-consistently, they can easily access to much larger cosmological volumes than hydrodynamical simulations.
In addition, a fast exploration of the parameter space and an efficient investigation of the influence of different specific assumptions are possible, thanks to the very limited computational costs.
We take advantage of the state-of-the-art semi-analytic model GAlaxy Evolution and Assembly (GAEA, \citet{DeLucia&Blaizot2007,DeLucia2014MNRAS,Hirschmann2016ADS}), coupled to large cosmological N-body simulations,
and analyse the properties of host DLA galaxies, as well as their connection with dark matter haloes. GAEA accounts for an explicit partition of the cold gas between atomic and molecular hydrogen, but assumes that all cold gas is associated with galaxy disks. Our approach therefore ignores the contribution to DLAs from filamentary structures or gas outflows, and tests to what extent current estimates of DLA statistics can be explained by the gas in galaxy disks. 

The specific questions that we will address in our study include:
\begin{itemize}
\item What is the typical virial mass of dark matter haloes that host DLAs?
\item To what extent can we reproduce the observed DLA statistical properties, by only considering the ISM associated with galaxies? 
\item What drives the evolution of $\Omega_{\rm DLA}$ with $z$, and what is the contribution to this quantity of galaxies with different mass?
\end{itemize}

The paper is organized as follows: in Section 2, we briefly present the semi-analytical model and the N-body simulations used in our study. We then discuss some basic predictions of our model, and the method that we have used to quantify the contribution of dark matter haloes that are not resolved by our simulations. In Section 3, we describe the methodology adopted to create our simulated sample of DLAs, and discuss model predictions in Section 4. In Section 5, we discuss our results in the framework of recent studies, and highlight model improvements/developments that could lead to a better agreement between model predictions and observational results. Finally, in Section 6, we give a summary of our results. 

\section{Properties of the simulated galaxies }
\label{sec:Properties-simulated-gals}
\subsection{The N-body Simulations}
The adopted physical model for the evolution of galaxies and their baryonic components is coupled to the output of cosmological dark matter simulations, as detailed in  \citet{DeLucia&Blaizot2007}. In this study, we use dark matter merger trees from two cosmological N-body simulations: the Millennium simulation (MSI; \citealt{Springel_MS2005}), and the Millennium II simulation (MSII; \citealt{BoylanMSII2009MNRAS}).

Both the MSI and the MSII assume a WMAP$1$ cosmology, with $\Omega_m = 0.25$, $\Omega_b= 0.045$, $\Omega_λ = 0.75$, $h = 0.73$ and $\sigma_8 = 0.9$. Recent measurements from Planck \citep{Planck2016A&A...594A...1P} and WMAP9  \citep{Bennet2013ApJ} provide slightly different cosmological parameters and, in particular, a larger value for $\Omega_m$ and a lower one for $\sigma_8$. As shown in previous work \citep{Wang2008JCAP, Guo2013MNRAS}, however, these differences are expected to have little influence on model predictions, once model parameters are tuned to reproduce a given set of observables in the local Universe.

\begin{figure}
\begin{center}
 \includegraphics[width=\columnwidth]{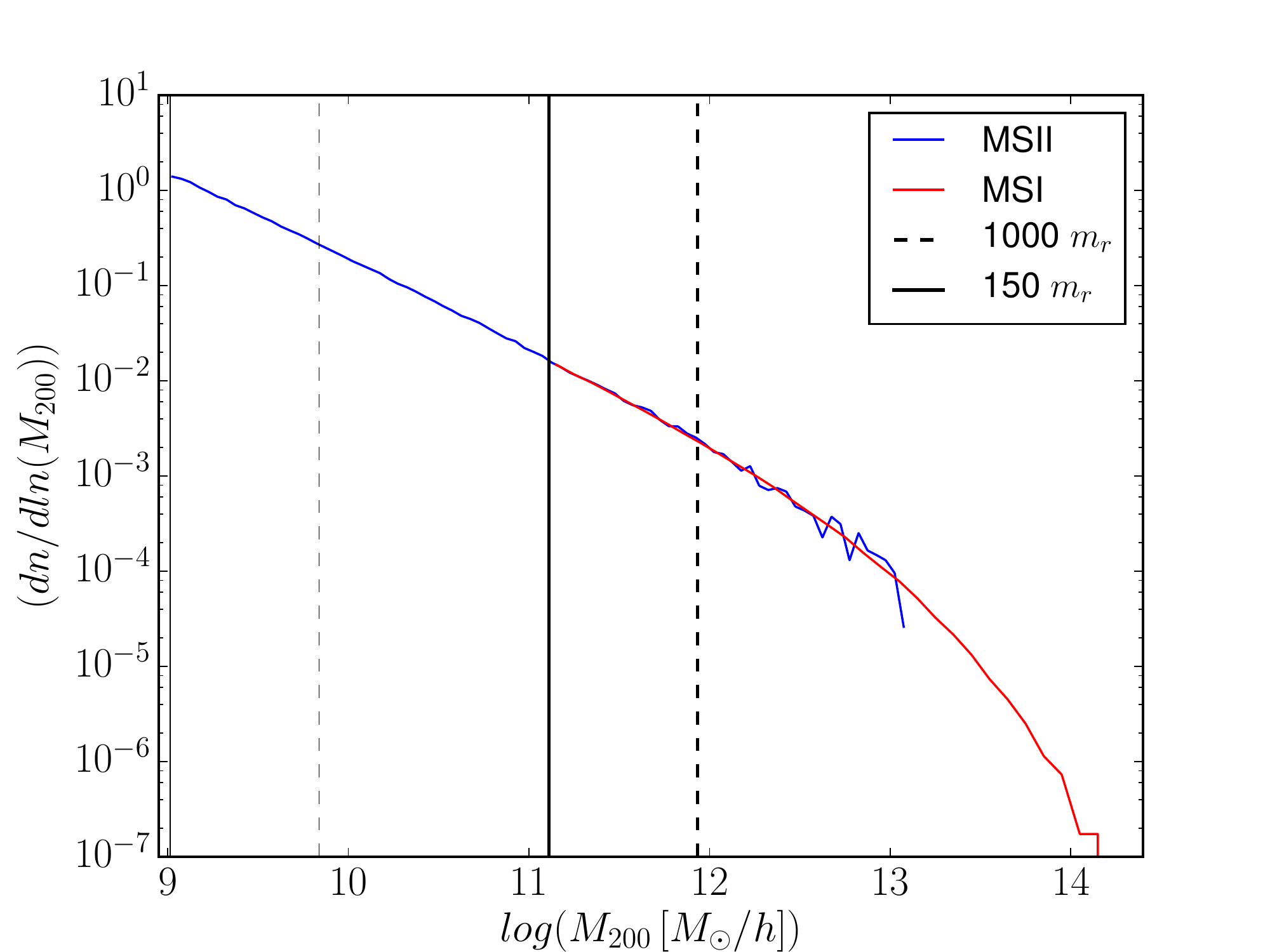}
 \caption{Comparison between the halo mass function, at $z=2$, derived  from the MSI and MSII (red and blue solid lines, respectively). The vertical solid and dashed lines correspond to $150$ times and $1000$ times, respectively, the DM particle mass for the two simulations. In our analysis, we will assume that haloes with more than 150 particles are well resolved in both simulations.}
   \label{fig:HMF-comparison-MSI-MSII}
\end{center}
\end{figure}

The particle mass is  $m_{DM}=8.61\times{10}^8 {M}_{\odot} h^{-1}$ for MSI and $m_{DM}=6.89 \, \times \, {10}^6 {M}_{\odot} h^{-1}$ for MSII, and the box size length
$L_{box}=500 {\rm c \,Mpc\, h}^{-1}$ and $L=100 {\rm c\, Mpc\, h}^{-1}$, respectively. 
In Fig.~\ref{fig:HMF-comparison-MSI-MSII}, we show the halo mass function (HMF) predicted from the two simulations at $z= 2$, where the halo mass is defined as the mass contained in a sphere which encloses an overdensity corresponding to 200 times the critical density of the Universe ($M_{200}$). In the following, we will consider as resolved all haloes that contain at least 150 particles. This corresponds to  $\sim {10}^{11} M_{\odot}/ h$ for the MSI and $\sim 10^9 M_{ \odot}/h$ for the MSII.
Below, we will combine the two simulations by selecting galaxies in haloes more massive than ${10}^{11.5} M_{\odot}$ from the MSI, and those residing in less massive haloes from the MSII. To investigate equal physical volumes in the MSI and MSII, we will subdivide the MSI box in 125 subboxes, with volume equal to that of the MSII.

\subsection{The semi-analytical model GAEA}
\label{sec:GAEA}
The GAlaxy Evolution and Assembly (GAEA) semi-analytic model, at the basis of this work, is an evolution of the model originally described in \citet{DeLucia&Blaizot2007}, with significant updates that have been published in the last years \citep[see, in particular,][]{DeLucia2014MNRAS,Hirschmann2016ADS}. In this study, we use the version of the model that includes an explicit treatment of the partition of cold gas in its atomic and molecular components \citep{Xie2017}. Specifically, we adopt the fiducial run presented in the work by \cite{Xie2017}, based on the empirical prescriptions by \citet{Blitz-Rosolowsky2006}. We will refer to this as the BR run in the following.

The GAEA model describes the evolution of four different baryonic reservoirs associated with a dark matter halo: (i) a hot gas reservoir that can grow due to cosmological accretion and stellar feedback, and from which gas cools onto the gaseous disks of central galaxies; (ii) a cold gas component associated with model galaxies from which stars form, and whose mass is affected by gas recycling due to stellar evolution and by stellar feedback; (iii) a stellar component for each model galaxy; and (iv) an ejected component that stores the gas that has been removed from the inter-stellar medium (ISM) of galaxies (i.e. cannot participate to star formation), and that can be later re-accreted onto the hot component associated with the parent dark matter halo. 

The BR prescription, described by \cite{Xie2017}, allows a  partition of the cold gas into atomic ($\rm HI$) and molecular (H$_2$) hydrogen, and has been tuned to reproduce the observed HI mass function at z=0. The ratio of molecular to atomic hydrogen,  $R_{\rm mol}=\Sigma_{H2} /\Sigma_{HI}$, depends on 4 physical properties of model galaxies: the mass of the {\it cold gas} ($M_{\rm CG}$, that in our model corresponds to gas with temperature below $10^4$~K), the galaxy stellar mass ($M_{\star}$), the size of the gaseous disc ($R_{\text{CG,d}}$ ), and the size of the stellar disc ($R_{\star,d}$). Using the empirical relation by \citet{Blitz-Rosolowsky2006}, the molecular fraction can be expressed as: 
\begin{displaymath}
R_{mol}= {\bigg(\frac{P_{ext}}{P_0} \bigg)}^{\alpha}
\end{displaymath}
where 
$P_0$ is the external pressure of molecular clumps and its logarithmic value is assumed to be $log(P_0/ k_B [{\rm {cm}^{-3} K}])=4.54$,
$\alpha = 0.92$, 
$P_{\text{ext}}=\frac{\pi}{2} G \Sigma_{\textsc{CG}} \big[ \Sigma_{\textsc{CG}} + f_{\sigma} \Sigma_{\star} \big]$,
$\Sigma_{\star}$ is the stellar surface density, and $\Sigma_{CG}$ is the cold gas surface density. The latter is  estimated in 21 logarithmic annuli (see original paper by \citealt{Xie2017} for details).

In our model, $R_{\text{CG,d}}$ and $R_{\star,d}$ are estimated from the specific angular momentum of the gaseous ($J_{\rm CG}$) and stellar ($J_{\star}$) disk component, respectively, assuming both are well described by an exponential profile: 
\begin{equation}
R_{\text{CG,d}}= \frac{J_{\rm CG} / M_{\rm CG} }{2 V_{max}}
\label{eq:scale-radius-gas}
\end{equation}

\begin{equation}
R_{\star,d}= \frac{J_{\star} / M_{\star} }{2 V_{max}} 
\label{eq:scale-radius-star}
\end{equation}
where $V_{max}$ is the maximum circular velocity of the dark matter halo.

\begin{figure}
\begin{center}
 \includegraphics[width=1.\columnwidth]{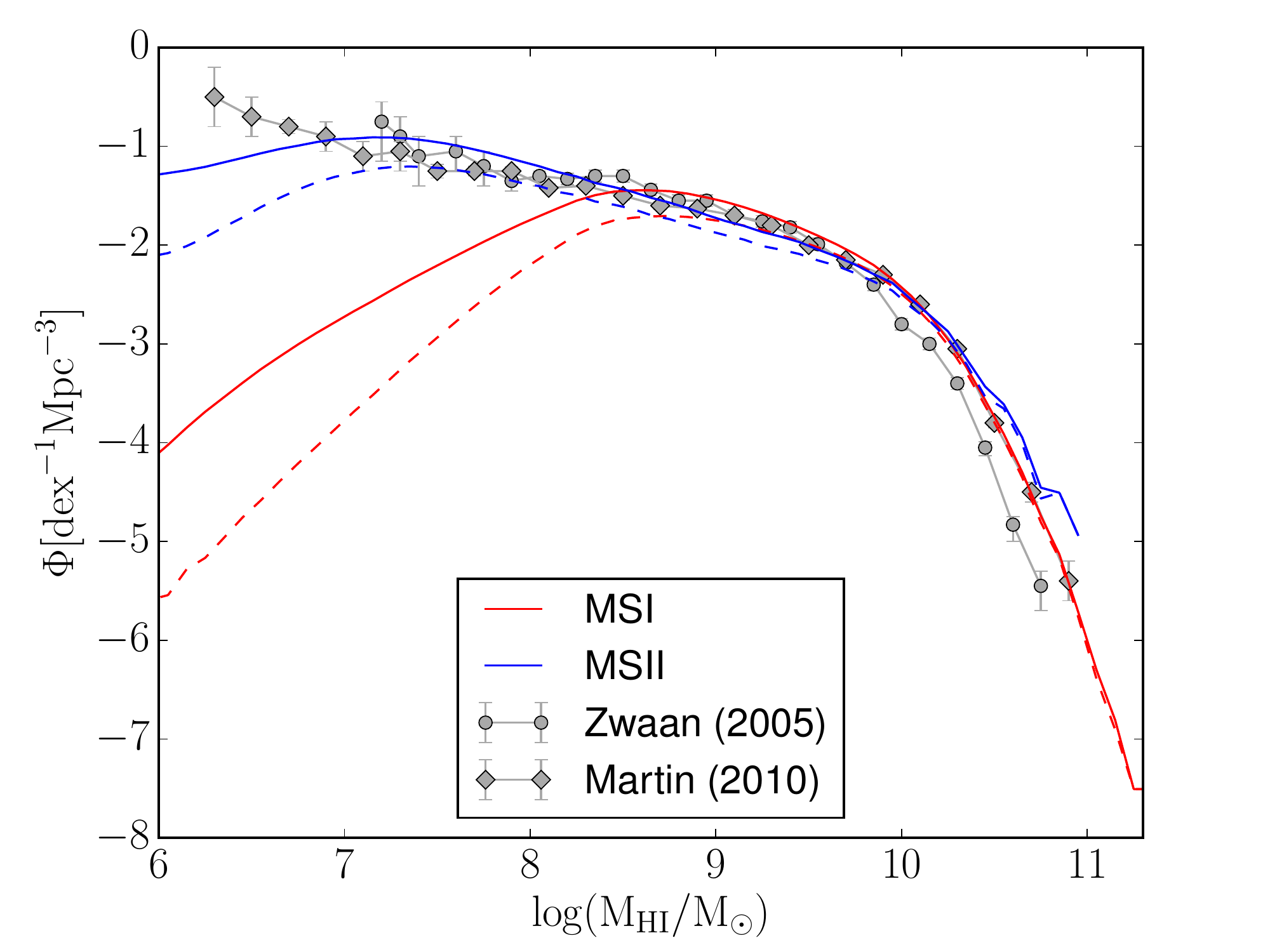}
 \caption{The HI mass function predicted for the MSI (red) and MSII (blue) simulations at $z=0$. Solid and dashed lines are used for all model galaxies and for centrals only, respectively. Dark grey symbols with error bars show the observational measurements by \citet{Zwaan2005MNRAS} and \citet{Martin2010HIMF}. These are based on the blind HI surveys HIPASS (\citealt{Meyer-HIPASS2004MNRAS}, limited to z < 0.04) and ALFALFA (\citealt{Giovanelli-ALFALFA2005AJ}, limited to z < 0.06), respectively. We apply the same stellar mass cuts adopted in \citet{Spinelli2019}.}
   \label{fig:HI-MF}
\end{center}
\end{figure}
Fig.~\ref{fig:HI-MF} shows the HI mass function predicted by GAEA at $z=0$, and compares model predictions with observational results by \citet{Zwaan2005MNRAS} and \cite{Martin2010HIMF}. The model runs used in this paper are based on the  Millennium I (red line) and Millennium II (blue line) simulations (see next section), that resolve DM haloes down to $\sim {10}^{11} {M}_{\odot}$ and $\sim {10}^9 {M}_{\odot}$ respectively. We consider as completeness limit for the cold gas mass at $z=0$ the values $M_{\rm CG} \sim {10}^8 {M}_{\odot}$ and $M_{\rm CG}\sim {10}^7 {M}_{\odot}$ for the MSI and MSII, respectively (for details, see \citealt{Spinelli2019}).

As mentioned above, the observed HI mass function in the local Universe has been used as the primary constraint for the BR model. Previous works have shown that the same model is able to reproduce a number of important additional observational constrains including scaling relations between the atomic/molecular mass and stellar mass, and the observed evolution of the mass-metallicity relation up to $z\sim 3$ \citep{Hirschmann2016ADS,Xie2017,Zoldan2017MNRAS}. This is relevant for our study that will include an analysis of the metallicities predicted for DLAs.

\subsection{HI cosmic density}

\begin{figure}
\begin{center}
 \hspace{-1.1 cm}
 \includegraphics[width=1.15\columnwidth]{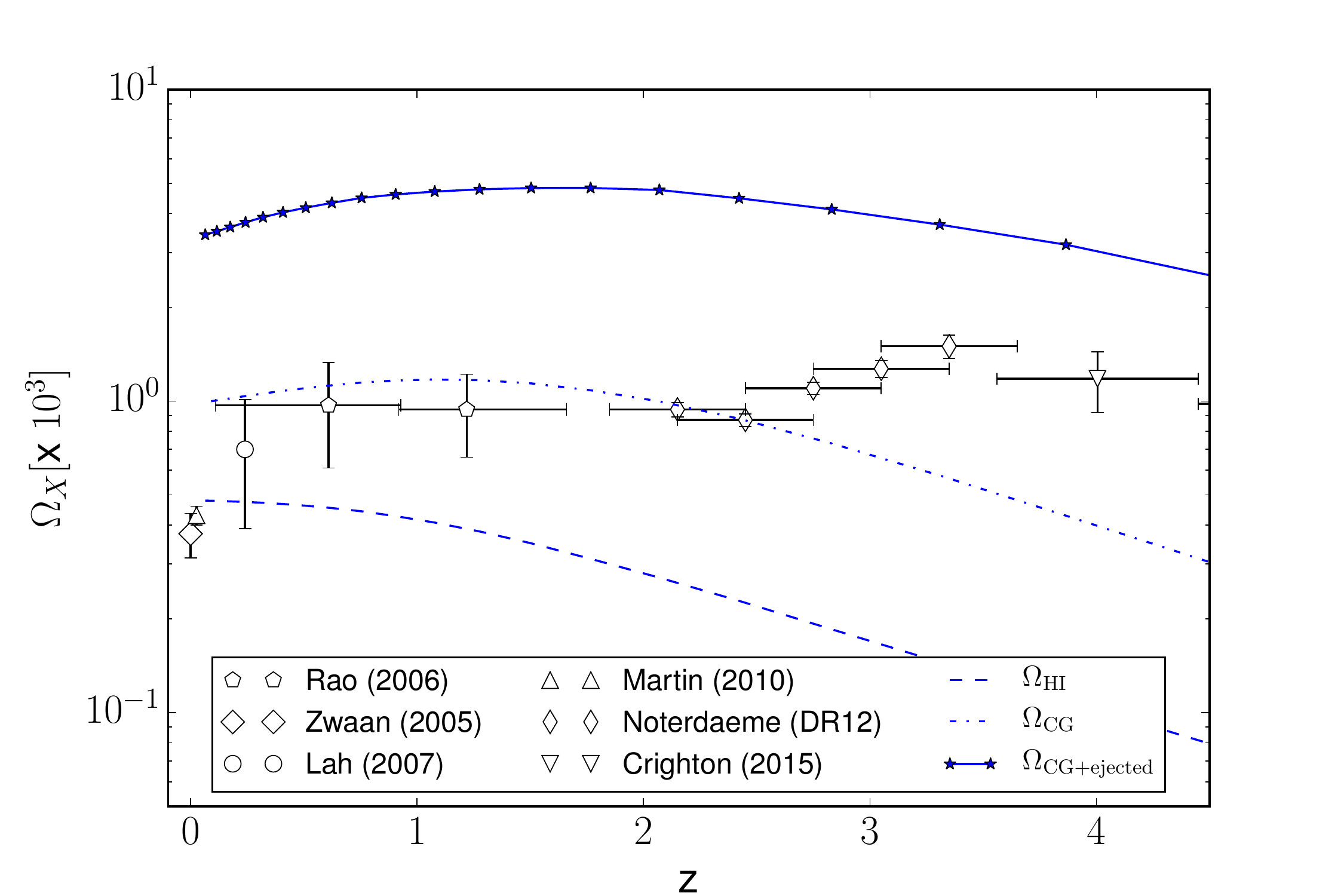}
 \caption{The comoving density evolution of the atomic hydrogen, cold and ejected gas (dashed, dot-dashed, and solid lines, respectively), obtained by summing the corresponding components of all model galaxies down to the resolution limits of the two simulations (see text for details). Model predictions are compared with observational measurements of $\Omega_{\rm HI}$ collected by \citet{Crighton2015}.}
  \label{fig:Omega_HI-evolution}
 \end{center}
 \end{figure}

We have estimated the comoving density of the atomic hydrogen, and cold and ejected gas ($\Omega_{\rm HI}$, $\Omega_{CG}$ and $\Omega_{ejected}$) in our simulated Universe, summing the corresponding gaseous components of all model galaxies residing in haloes above our adopted resolution limits (for each observable component X, $\Omega_{\rm X} (z)= \frac{{\rho}_{x} \, (z)}{\rho_{c}(0)}$). In particular, we have summed the comoving gas density measured in the MSII box, considering haloes in the mass range ${10}^{9.2} \leq M_{200} < {10}^{11.5} M_{\odot}$ to the average comoving gas density measured  from haloes with $M_{200} \geq {10}^{11.5} M_{\odot}$ in the 125 sub-boxes of the MSI, each with a volume equal to the volume of the MSII box. Fig.~\ref{fig:Omega_HI-evolution} shows these model predictions together with observational measurements of $\Omega_{\rm HI}$ from \cite{Crighton2015}. In this figure, we have corrected for the critical density value corresponding to the cosmology adopted by \cite{Crighton2015}.

Our simulated estimate of $\Omega_{HI}$ is a factor $\sim 2.5$ below the observational estimates based on DLA surveys up to $z \sim 2$, and further decreases at higher redshift. The low z behaviour of the predicted $\Omega_{HI}$ is not surprising, because the GAEA model is tuned to reproduce the HI mass function observed in the local Universe by \cite{Martin2010HIMF}, whose integrated value is a factor $\sim 2$ lower than the HI cosmic density estimate by \citet{Lah2007MNRAS} at $z\sim 0.24$ (also based on emission lines measurements), and than estimates based on statistical analysis of DLAs at higher redshift \citep{Rao06,Noterdaeme2012}.

The decrease of $\Omega_{HI}$ at high redshift ($z>3$) is more difficult to explain.
\cite{Spinelli2019} show that the largest contribution to $\Omega_{HI}$ in our model is given by haloes with mass ${10}^{10} M_{\odot} \leq M_{200} \leq {10}^{12} M_{\odot}$, and that $\Omega_{HI}$ decreases with increasing redshift for more massive haloes while it flattens for less massive haloes.
The decrease of $\Omega_{HI}$ at higher redshift is found also for independent semi-analytical models that consider a similar mass range of dark matter haloes contributing to the HI density \citep[e.g.][]{Lagos2011MNRAS, Berry2014MNRAS}.

A possible solution to this problem is to increase the contribution of intermediate and low-mass haloes to $\Omega_{HI}$ at high redshift, that we may be underestimating because of the adopted physical prescriptions and resolution limits of our simulations. It is difficult to quantify precisely the impact of resolution on our results, as it can affect both the missing HI content of the unresolved isolated haloes and the HI content of the satellite galaxies hosted in the resolved haloes. Considering the resolution limit of the MSII, and the observed scaling relation between the HI to stellar mass ratio and galaxy stellar mass, we expect that the largest contribution should come from the HI content of unresolved haloes. In the next section, we explain how we compute an estimate of such a contribution.

\subsection{Minimal HOD model}
\label{sec:Minimal_HOD_model}

To quantify whether low-mass haloes (i.e. below $\sim 10^{9} M_{\odot}$) can significantly increase the HI density in our simulated Universe, we populated the MSII box with haloes below its resolution using a simple halo occupation distribution (HOD model - see \citealt{Berlind2002ApJ} for an historical review). 

\begin{figure}
\begin{center}
 \hspace{+1.0 cm}
 \includegraphics[width=\columnwidth]{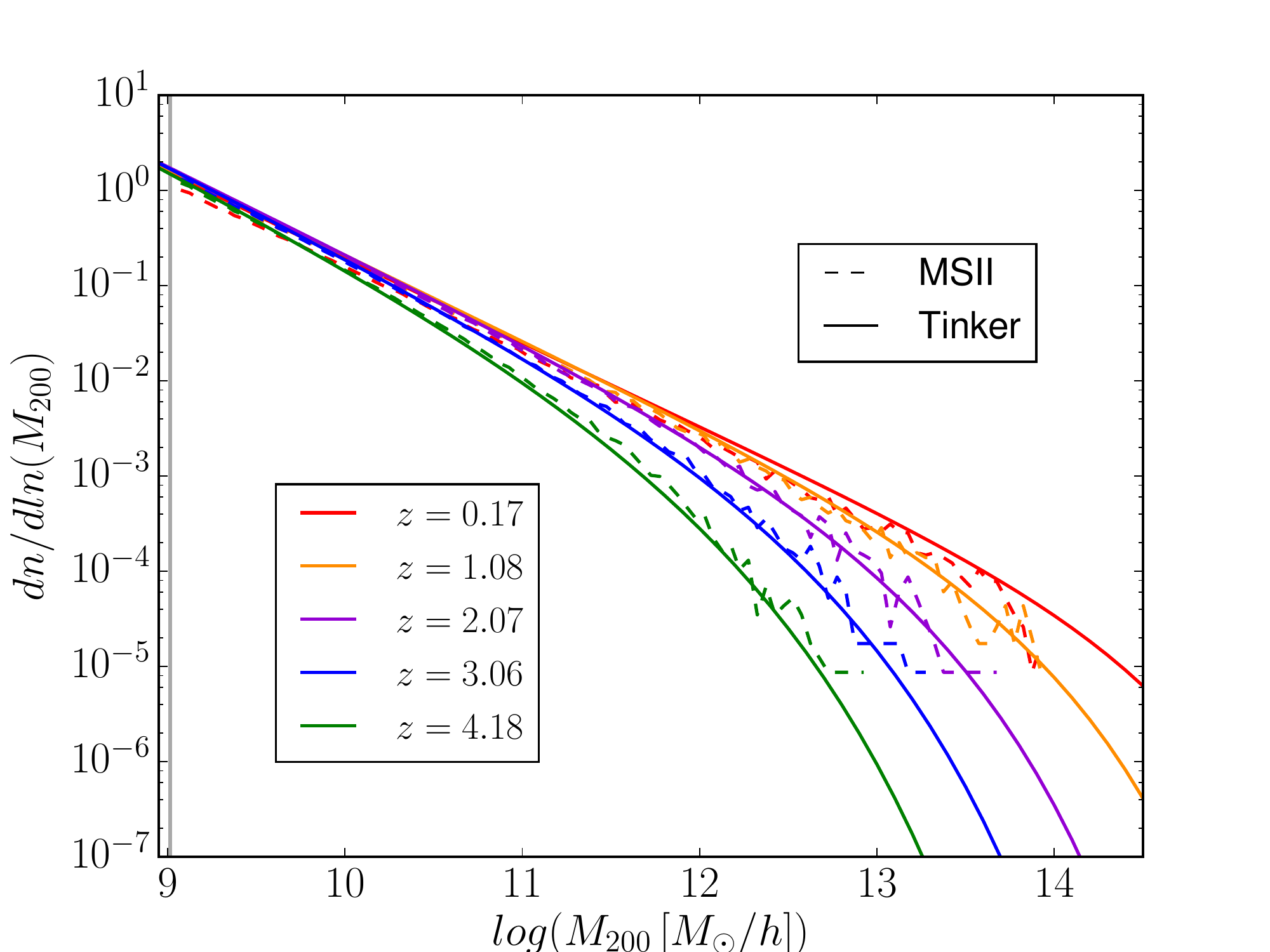}
 \caption{Comparison between the Tinker halo mass function (solid lines) and that estimated from the MSII (dashed lines), at 5 different redshifts, listed in the legend.The vertical line marks the resolution limit of the MSII simulation.}
\label{fig:HMF}
\end{center}
\end{figure}

The number of low-mass haloes to be added, and their mass distribution, have been derived integrating the HMF by \citet{Tinker2008} in the range $10^8 M_{\odot} \leq M_{200}<10^{9.2} M_{\odot} $, and using the cosmological parameters adopted for the Millennium simulations. We have checked that the shape and normalization of the Tinker HMF are consistent with those derived from the MSII and MSI. 
This can be appreciated in Fig.~\ref{fig:HMF}, where we compare the Tinker HMF (solid lines) with that measured from the MSII (dashed lines), at 5 different redshifts.

\begin{figure*}
\begin{center}
 \includegraphics[width=2.2\columnwidth]{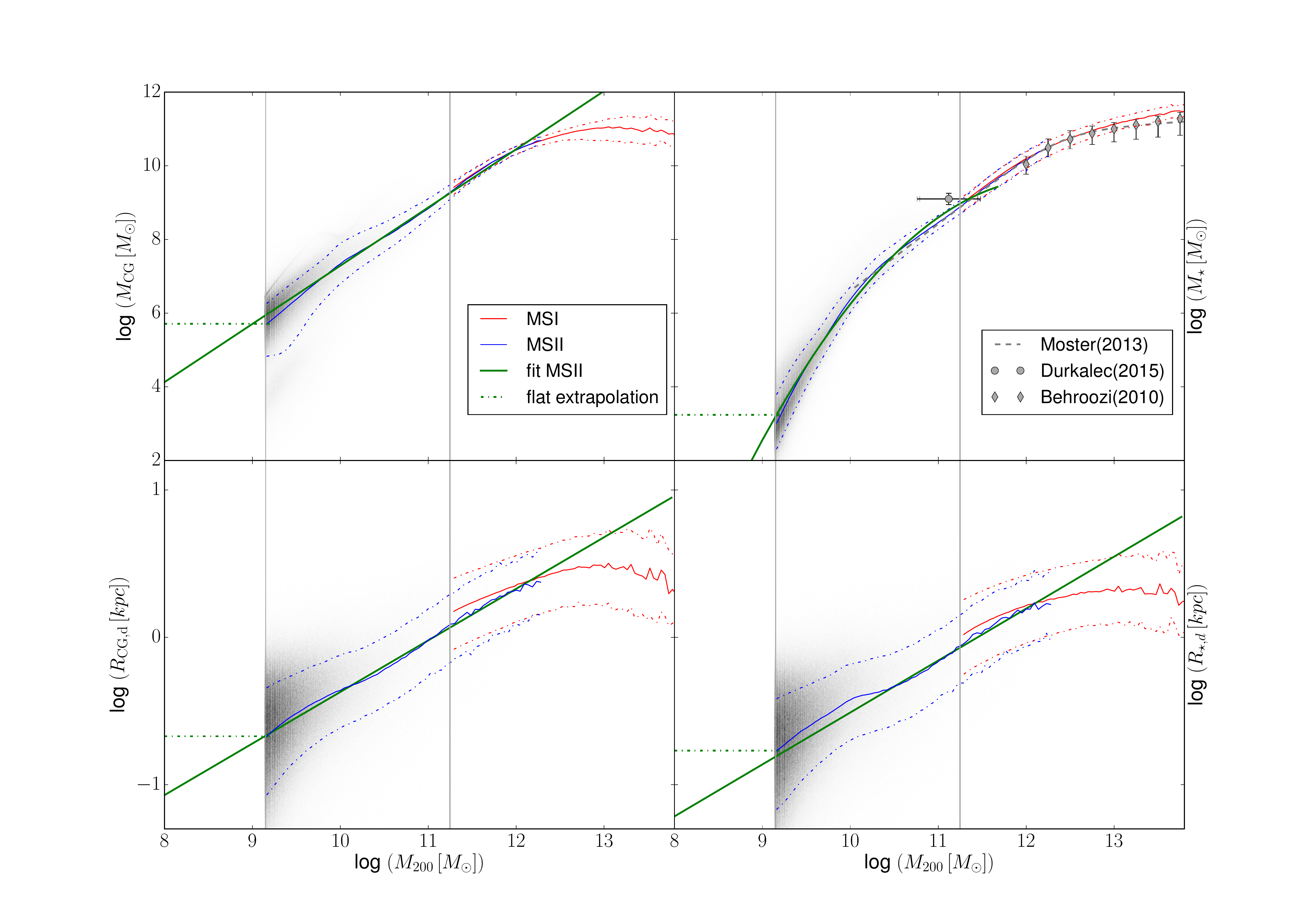}
 \caption{Scaling relation for $M_{\rm CG}$, $M_{\star}$, $R_{{\rm CG},d}$, and $R_{\star,d}$ as a function of halo mass for central galaxies in MSI (red) and MSII (blue), at redshift $z=2.07$. The grey color gradient highlights the number density of MSII central galaxies. The solid and dot-dashed lines show respectively the median and percentiles (16th and 84th) of the distributions. The solid green lines show the linear fit for all relations extracted from the MSII, except for the $M_{\star}$ vs $M_{200}$ relation. In this case, we use a polynomial fit of second order. The dot-dashed lines are flat extrapolations of the scaling relations, normalized to median values corresponding to the lowest halo mass bin resolved in the MSII. In each panel, the vertical solid thick (thin) line shows the resolution limit of the MSI (MSII). For the SMHM relation we show also observational estimates \citep{Behroozi2010ApJ,Durkalec2015A&A} and the fitting function derived by \citet{Moster2013MNRAS} for central galaxies.} 
\label{fig:scaling-relations}
\end{center}
\end{figure*}

Considering the low mass of the haloes treated with the HOD model, we have populated them only with central galaxies, since we do not expect that they host satellites, and distributed them at random positions inside the MSII box. We assign 5 physical quantities to galaxies in the HOD model: stellar mass($M_{\star}$), cold gas mass ($M_{\rm CG}$), scale radius of the gaseous disk ($R_{{\rm CG},d}$), scale radius of the stellar disk ($R_{\star,d}$), and abundance ratio $[\rm Fe/H]$. These quantities are derived extrapolating the scaling relations obtained from our semi-analytic model run on the MSI and MSII. The scaling relations for the first four quantities are shown in Fig.~\ref{fig:scaling-relations} for $z=2$ (these scaling relations evolve slowly as a function of redshift), while the extrapolation of $[\rm Fe/H]$ is treated in detail in subsection \ref{Sect:Metallicity}. 
In the mass regime where MSI and MSII overlap we observe a nice convergence of the scaling relations 
(e.g. the difference between galaxy stellar mass of MSI and MSII is less than 10 $\%$ for haloes with $M_{200} \sim {10}^{11} M_{\odot}$ ). 
Fig. ~\ref{fig:scaling-relations} also shows that the predicted SMHM relation is in good agreement with the observational estimates \citep{Behroozi2010ApJ,Moster2013MNRAS,Durkalec2015A&A}.

We have considered two different extrapolations of the predicted scaling relations at each of the snapshots analysed: (i) a linear fit of the median relation obtained for galaxies in MSII (a second order polynomial for the stellar mass - halo mass relation); (ii) a flat extrapolation normalized to the value obtained for the smallest haloes in the MSII. 

\begin{figure}
\begin{center}
\hspace*{-0.5 cm}
\includegraphics[width=1.1\columnwidth]{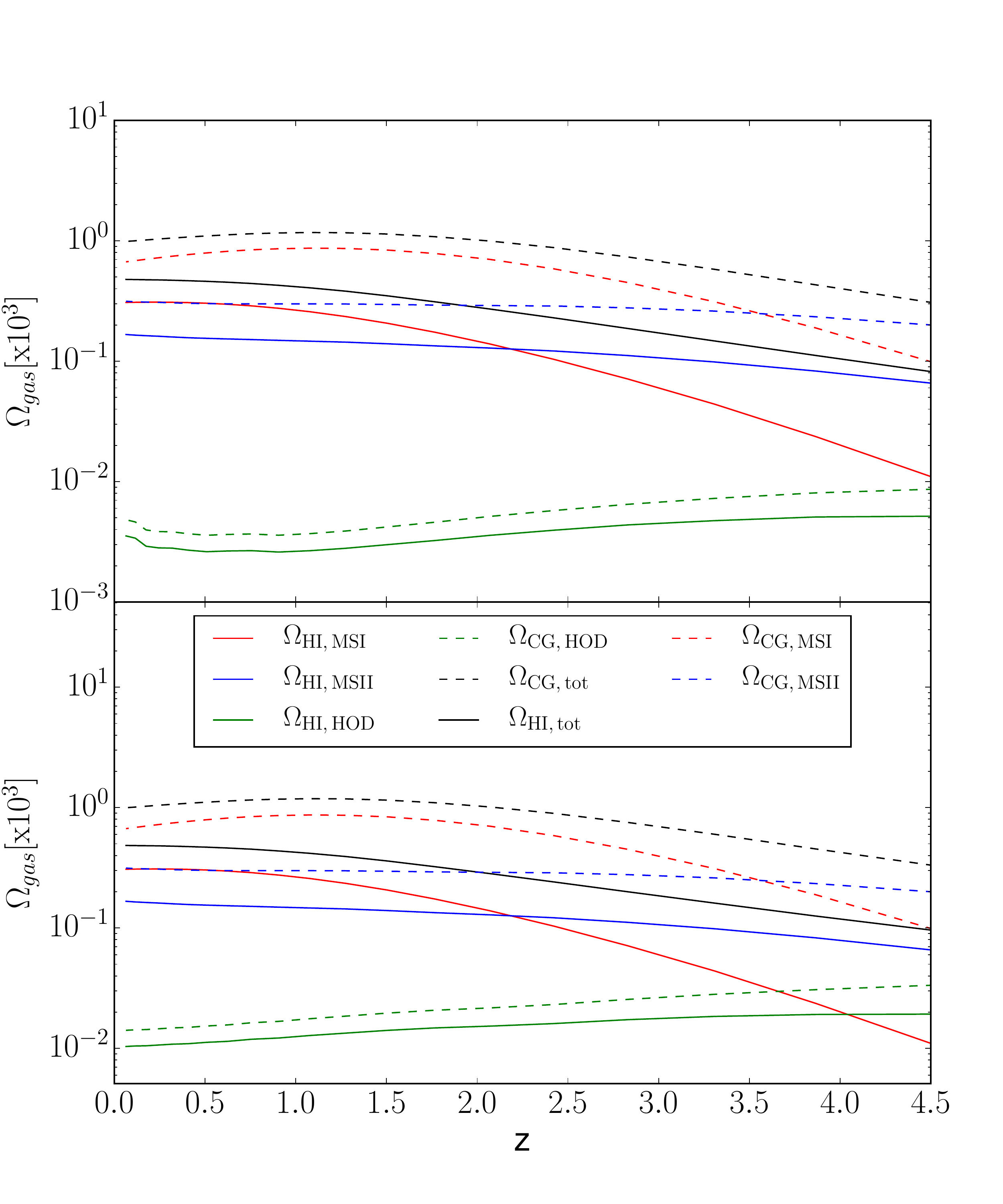}
\caption{Evolution of the comoving density of atomic hydrogen and cold gas content of model galaxies (HI solid, CG dashed), residing in massive haloes of MSI (red), intermediate-mass haloes of MSII (blue) and low-mass haloes of our HOD extension (green). The black solid (dashed) line shows the total content of HI (CG) in our simulated universe.
The top panel shows the contribution coming from HOD galaxies when considering the linear (or 2nd-order for the stellar mass - halo mass relation) extrapolation of the scaling relations obtained from the MSII galaxies, while the bottom panel corresponds to the flat extrapolation of the scaling relations.}
  \label{fig:Omega_gas_in_MSI_MSII_plus_HOD}
 \end{center}
 \end{figure}

Based on the values extrapolated for $M_{\star}$, $M_{\rm CG}$, $R_{\star,d}$, and $R_{{\rm CG},d}$, we then estimate the molecular fraction using the empirical relation by \citet{Blitz-Rosolowsky2006} in 21 annuli. For each galaxy in the HOD catalogue, we store the integrated molecular gas fraction in the disk ($R_{mol}$), that we use to estimate the atomic gas mass. Fig. \ref{fig:Omega_gas_in_MSI_MSII_plus_HOD} shows the evolution as a function of redshift of the comoving density of HI and cold gas (solid and dashed lines, respectively) computed considering all DM haloes from the MSI, MSI and HOD model. The contribution to $\Omega_{\rm CG}$ coming from the HOD galaxies becomes non-negligible only at relatively high redshifts (e.g. for $z> 4.5$), and only when considering a flat extrapolation of the scaling relations. The contribution to the cosmic density of neutral hydrogen is dominated by MSI haloes up to redshift $z \simeq 2.3$, when the MSII starts dominating. For the cold gas, the cross-over between the MSI and MSII takes place at $z \gtrsim 3.5$. 

We have studied the effect of different halo mass cuts on the cosmic HI content in our simulated Universe, finding little differences. 
In the following, we adopt the following fiducial cuts: we select haloes with log$(\frac{M_{200}}{ M_{\odot}})$ in the range $[8,9.2)$ from the HOD, haloes with log$(\frac{M_{200}}{ M_{\odot}})$ in the range $[11.5,max)$ from MSI, and haloes from the MSII in the intermediate regime.

\section{Simulated DLA catalogs}
\label{Sect:CatalogCreation}

In order to produce samples of simulated DLAs to be compared with observational data we have thrown random lines of sight (LOS) in the volume of the composite simulation described in the previous section. 
To cover a large halo mass range $({10}^8-{10}^{15} M_{\odot} )$, we consider together the galaxies hosted in the DM haloes selected from the MSI, MSII and in those added using the HOD, according to the halo mass cuts described in the previous section. The physical properties assigned to HOD galaxies are derived adopting the flat extrapolation.

We have subdivided the volume of MSI in 125 sub-boxes, of volume equal to that of the MSII box ($L_{box}=100 h^{-1} \rm Mpc$ ), and constructed at each redshift analysed 125 realizations that differ only for the MSI contribution. In this way, it is possible to investigate the impact of the cosmic variance on the DLA observables considered in this study, for the DM haloes that are well resolved in the MSI (i.e. with $M_{200}> {10}^{11} M_{\odot}$). Since cosmic variance is more important for rarer (i.e. more massive) systems, we expect that it does not play an important role for the intermediate mass haloes that are selected from the MSII simulation. 

For each simulation snapshot in the redshift range of interest, 
we throw 100,000 random LOS, parallel to the $z-$axis, for each of the 125 realizations considered. This provides us with 125 simulated DLA catalogs.

\subsection{$\rm N_{HI}$ estimate}

For each galactic disk, we assume that the gas density profile follows a double-exponential profile\footnote{We have tested that assuming an isothermal vertical profile for the gas in the ISM \citep{VanDerKruit2011}, or alternative vertical profiles suggested in the literature (see Appendix B), does not affect significantly our results.}:
\begin{equation}
    \rho_{\rm CG}(r,z)=\rho_0 ~ e^{-{r}/{R_{{\rm CG},d}}} ~e^{-{z}/{z_0}}
\label{eq:CG density profile}
\end{equation}
where $\rho_0$ is the normalization of the 3D density profile for the gaseous disk, $R_{{\rm CG},d}$ and $z_0$ are the scale-radius and the scale-height of the gaseous disk, respectively. 

For the scale-height parameter $z_0$, we apply a linear dependence on the scale radius:
\begin{equation}
z_0=\frac{R_{{\rm CG},d}}{A},
\label{eq:rel-Rs-z0}
\end{equation}
and test two different values of the fudge factor A: $=7.3$ and $=4$. The former choice relies on observational relations
valid for stellar disks in the local Universe \citep{Kregel2002MNRAS}, while the latter is motivated by observations of thicker galactic stellar disks at $z \sim 2$ \citep{Elmegreen2017ApJ}.

The HI density profile can be written as:
\begin{equation}
  \rho_{\rm HI}(r,z)=(1- f_{mol}(r) \, ) \, \rho_{\rm CG}(r,z)  
\label{HI density profile}
\end{equation}
where the molecular fraction $f_{mol}$ has been estimated using the BR prescription (described in Section \ref{sec:GAEA}), in 21 logarithmic radial bins between $r=0$ and $r=10 \, R_{{\rm CG},d}$. When a given LOS intersects a galaxy with a distance (impact parameter) $b \leq 10 R_{{\rm CG},d}$, the hydrogen column density ($\rm N_{ HI}$) contributed by the galaxy can be estimated by integrating the HI density profile along the LOS. The value of $\rm N_{ HI}$ depends then on the impact parameter and on the inclination of the galactic plane with respect to the LOS.

The assumption that cold gas in model disk galaxies is distributed according to an exponential density profile is in good agreement with observational findings \citep[e.g.][]{Wang2014MNRAS}. 

We have also considered the contributions from close galaxies/pairs to each absorption feature. Adopting a FoF-like merging algorithm, we summed all column densities of absorbing systems, intersected by the same LOS, with a maximum velocity offset of $\Delta v \leq 2000 ~{km}/{s}$. 
Our merging algorithm works as follows: we firstly subdivide the systems along the same LOS into groups of close systems sorted along the z-coordinate. Then, we merge the two nearest systems in each group, estimate the barycentre of the pair and re-estimated the distance between the first merged system and the other systems in each group. If necessary, we repeat the merging process and re-iterate until there is no other pair to merge. 

The estimated fraction of DLAs originated from multiple systems is large (more than $70 \%$ at $z=2$ for the $2M-2R$ model and more than $50 \%$ for the fiducial model). However, in most cases one single galaxy contributes significantly more than the others. 
In particular, if we consider only systems with column density $N_{\rm HI}\geq {10}^{17}  \, atoms \, {\rm cm}^{-2}$, in $87\%$ ($84\%$) of the cases more than $80 \%$ of the total hydrogen column density comes from one single galaxy while the cases where the contribution of each single galaxy is less than $50\%$ represent only $1\%$ ($0.5\%$) of the all cases for the $2M-2R$ model (for the fiducial model). Therefore the distribution of simulated DLA column densities is not significantly affected by the blending of close absorption features.

\subsection{Distribution of impact parameters versus $\rm N_{HI}$}

\begin{figure}
\begin{center}
\hspace*{-0.5 cm}
 \includegraphics[width=1.1\columnwidth]{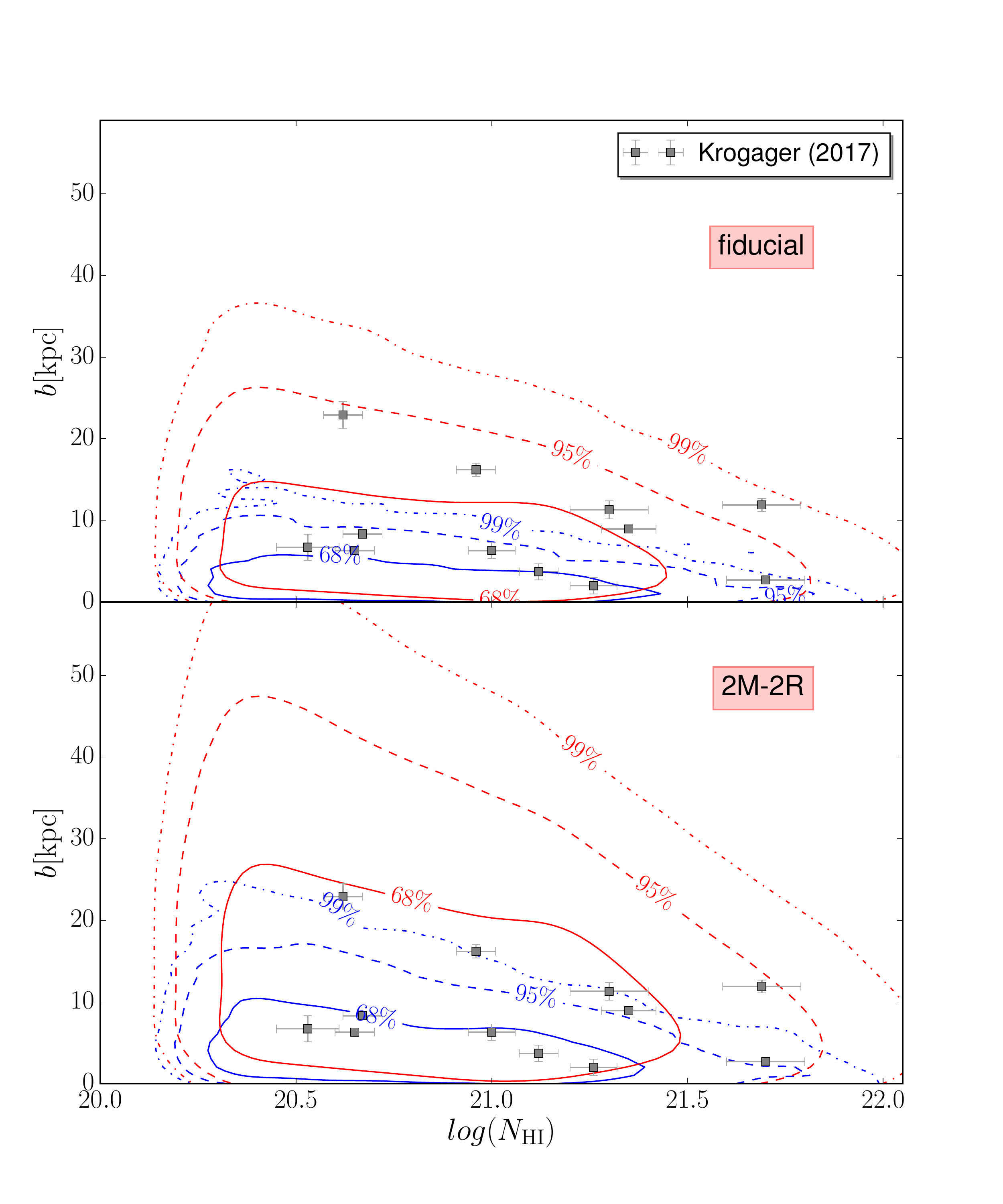}
 \caption{Impact parameter (b) as a function of the hydrogen column density $\rm N_{HI}$, for simulated DLAs in the redshift range $2<z<3$, compared to the observations by \citet[][grey symbols]{Krogager2017}. The top panel shows the contour distributions based on the fiducial model, while the bottom panel shows results obtained multiplying by a factor 2 both the scale radius and the cold gas mass of all model galaxies ($2M-2R$ model, hereafter). Red and blue lines refer to galaxies in the MSI and MSII, respectively, and show different contour levels of the distribution as indicated in the legend. We apply a cut to the metallicity of model DLAs equal to $\rm [Fe/H]> -2.0$, for consistency with the observational measurements considered.}
   \label{fig:NHI-b}
\end{center}
\end{figure}
For our model galaxies, we define the impact parameter, $b$, as the distance between the LOS and the center of mass of the galaxy hosting the DLA. In observations, $b$ measures the projected distance between the (luminosity) center of the galaxy and the quasar sight-line piercing the cold gas. 

In Fig.~\ref{fig:NHI-b}, we show the distribution of impact parameters as a function of the hydrogen column density ($\rm N_{HI}$), obtained considering DLAs originating from the MSI and MSII haloes, in the redshift range $2<z<3$ and with $\rm [Fe/H]> -2$. Model predictions are compared with observational measurements by \citet{Krogager2017}, that cover the same redshift and metallicity range. The data point come partly from the literature and partly from an X-shooter follow-up campaign. The latter is the first sample of DLA counterparts at high redshift associated with a relatively high detection rate ($\sim 64 \%$), likely due to the adopted DLA pre-selection: $\rm EW_{SiII}> 1$\AA ($\rm EW_{SiII}$: rest-frame equivalent width of the $\rm Si_{II}$ line, with $\lambda = 1526$\AA) implying large metallicities. 

The top panel of Fig.~\ref{fig:NHI-b} shows results from the run of GAEA described in \ref{sec:GAEA}, our fiducial model, while the bottom panel shows results obtained multiplying by a factor 2 both the scale radius and the cold gas mass of all model galaxies. In the following, we will refer to this as the $2M-2R$ model. 

The largest $99$ per cent contour level of the simulated distribution, in both the fiducial and the $2M-2R$ model, encloses all the observed data. In the $2M-2R$ model, all data points fall inside the $95$ per cent contour level of the simulated distribution (for MSI haloes), and there is a more clear anti-correlation between impact parameters and column density. For the fiducial model, we find ${<b>}^{sim}_{DLA}=8.23$ ($3.00$) for MSI (MSII).
The corresponding values for the 2M-2R model are  ${<b>}^{sim}_{DLA}=14.63$ ($5.14$). The different mean value of the impact parameters and the different contour levels between DLA originated from the MSI and MSII haloes reflect the dependence of the galactic disk size on the virial radius of the halo where the galaxy resides.

Averaging and weighting over the relative contribution of MSI and MSII, we obtain ${<b>}^{sim}_{DLA}=5.53$ and ${<b>}^{sim}_{DLA}=10.03$ for the fiducial and the $2M-2R$ model, respectively. Both estimates are in agreement with the one found by \cite{Krogager2017}, ${<b>}_{DLA}=8.32$, with a slight preference for the  $2M-2R$ model. 

Extending the sample of observed DLAs towards lower redshift, \citet{RhodinC2018} reports ${<b>}_{DLA}=11.1$ kpc. Older work by \cite{Rao2011MNRAS} based on low-redshift DLAs counterparts found ${<b>}_{DLA}=17.4$ kpc, considering a larger metallicity cut ($\rm [Fe/H]> -1$). The different observational estimates depend on the adopted DLA pre-selections, on the techniques used to search for DLAs counterparts, and in part also on the expected redshift evolution of galaxy sizes that implies an evolution of the observed range of impact parameters.

It is important to bear in mind that all observations of DLA counterparts are biased against smaller impact parameters, for which it is difficult to detect the DLA counterparts \citep[as discussed in ][]{Krogager2017}. The technique adopted by \citet{Krogager2017} likely misses also some counterparts at large impact parameters due to the partial coverage of the FoV by the three long-slits while DLA systems with very high metallicities are not completely detected because of the dust bias \citep{Khare2012MNRAS}, which affects the colour selection of QSOs. Therefore, the comparison shown in Fig.~\ref{fig:NHI-b} should be considered more as qualitative than rigorous.


\subsection{Assigning metallicity to DLAs}
\label{Sect:Metallicity}

The GAEA model adopts a detailed chemical enrichment scheme that accounts for the finite lifetime of stars and the non-instantaneous recycling of metals, gas, and energy  \citep[]{DeLucia2014MNRAS}.

As discussed in previous work, the fiducial model used here is able to reproduce the observed evolution of the correlation between galaxy stellar mass and cold gas metallicity, up to $z\sim 2$ \citep{Hirschmann2016ADS,Xie2017}. This is an important achievement, met by only a few recently published theoretical models \citep[see discussion in ][]{Somerville-Dave2015ARA}. Our study offers an additional test to the model.  

As commonly done in DLA studies, we use the iron over hydrogen abundance ratio, $\big[\frac{\rm Fe}{\rm H}\big]$, as a proxy for the metallicity of the gaseous disks of our simulated galaxies. GAEA assumes a uniform distribution of the metals in the different baryonic components. So we can write:
\begin{equation}
\bigg[\frac{\rm Fe}{\rm H}\bigg]= \log \bigg( \frac{{\rm M}_{\rm Fe,d} \, \mu_{\rm H}}{{\rm M}_{\rm HI,d} \, \mu_{\rm Fe}}\bigg)- \log {\bigg( \frac{{\rm M}_{\rm Fe} \, \mu_{\rm H}}{{\rm M}_{\rm HI} \, \mu_{\rm Fe}}\bigg)}_{\odot}
\label{eq:uniform abundance ratio}
\end{equation}
where $\rm M_{Fe,d}$ and $\rm M_{HI,d}$ are the masses of Fe and HI in the cold gaseous disk of each galaxy, while $\mu_{\rm Fe}$ and $\mu_{\rm HI}$ are the corresponding mean atomic weights. ${[\rm Fe/H]}_{\odot}$ is the solar abundance ratio, that we take from \citet{Asplund2009ARA}.

\begin{figure}
\begin{center}
\hspace*{-0.8 cm}
 \includegraphics[width=1.3\columnwidth]{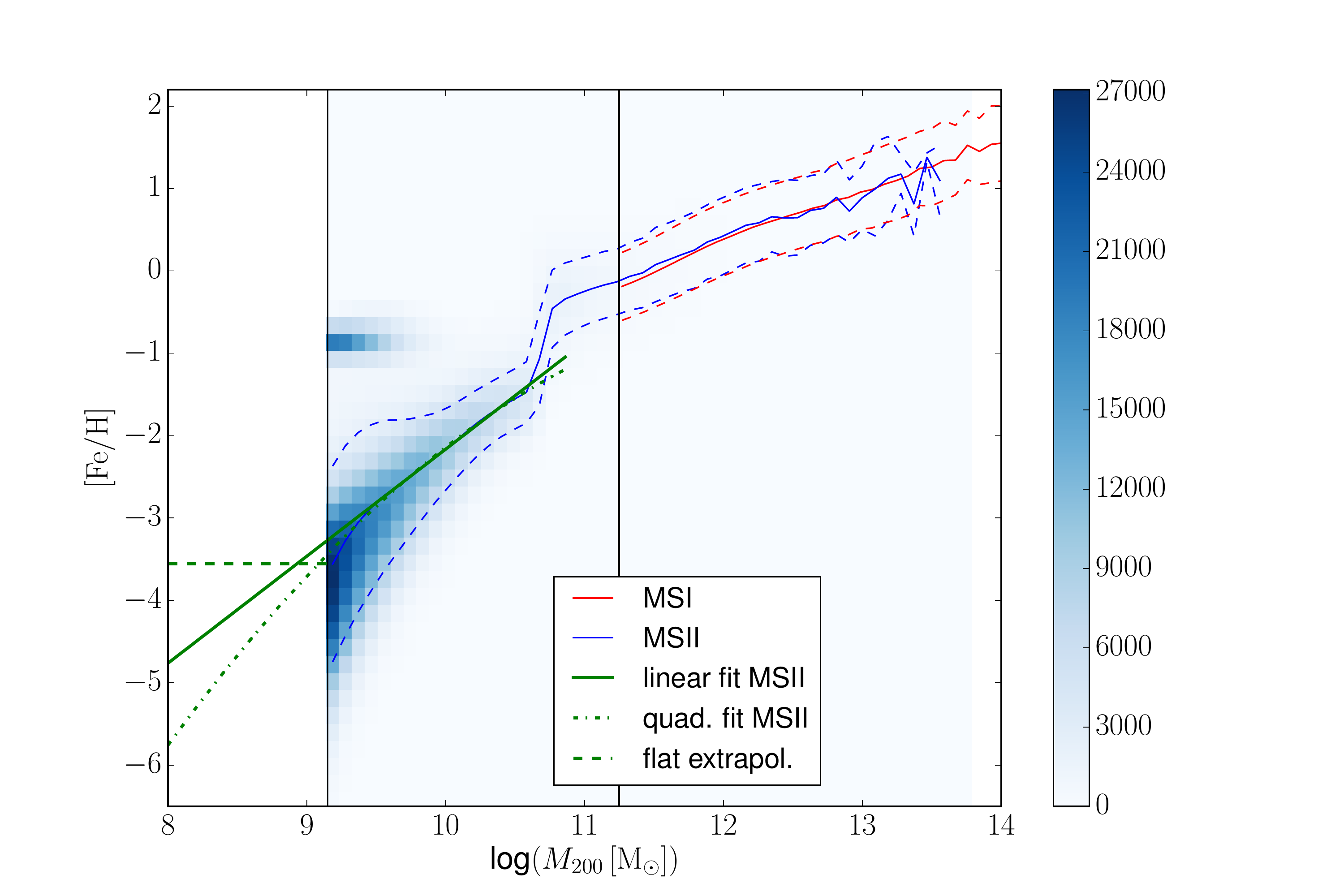}
 \caption{ $\rm [Fe/ H]$ as function of $log(M_{200})$ for central galaxies, at $z=2$. The solid blue (red) line shows the mean relation for MSII (MSI) galaxies. 
 The green solid, dashed and dot-dashed lines show the extrapolated linear fit to the mean relation measured for the MSII, a flat extrapolation, and an extrapolation based on a quadratic fit to the MSII results. The color coding quantifies the number density of the MSII central galaxies. The vertical lines show the resolution limits of the MSI and MSII.}
   \label{fig:FeH_vs_Mh}
\end{center}
\end{figure}

Fig.~\ref{fig:FeH_vs_Mh} shows the relation between the abundance ratio $[\rm Fe/H]$ and $log(\rm M_{200})$, for the central galaxies in the MSI and MSII. There is a good convergence between MSI and MSII in the galaxy mass range  ${10}^{8.5}<M_{\star}<{10}^{10} M_{\odot}$. For the extrapolation of the $[\rm Fe/H]-M_{200}$ relation to galaxies inside the haloes sampled by the HOD, we have used a linear regression in the mass range $\rm {10}^{9.2}<M_{200}<{10}^{10.6} [M_{\odot}]$, i.e. after the step-like feature visible in the figure.
This feature arises mainly as a consequence of a specific assumption of our galaxy formation model: for $\rm M_{200} \leq 5 \times {10}^{10} M_{\odot}$,  $95 \%$ of the new metals are ejected directly into the hot phase, instead of being mixed with the cold-gas in the ISM, as assumed for more massive haloes. This assumption was motivated by results from hydrodynamical simulations \citep{MacLow-Ferrara1999ApJ} and was helpful, in previous versions of our models, to reproduce the metal content of satellites in Milky-Way-like haloes \citep{Li2010MNRAS}. 

Considering the large $1-\sigma$ scatter of the predicted relation, we have applied it to the extrapolated values of the abundance ratio for HOD galaxies.
Once assigned the atomic hydrogen mass to HOD galaxies (following the procedure described in the previous section), we can use the extrapolated abundance ratio to assign an iron mass to each HOD galaxy.

We have also considered the effect due to the presence of a metallicity radial gradient. 
Specifically, we have assumed a slope consistent with the observational study by \citet{Christensen2014MR}:
\begin{equation}
    \Gamma= -0.022 ~ \text{dex} {\text{ kpc}}^{-1} 
	\label{eq:slope-gradient}
\end{equation}

\cite{Christensen2014MR} and \cite{RhodinC2018} found an almost universal metallicity gradient for a sample of DLAs observed in the redshift range $(2-3.5)$. \cite{Stott2014MNRAS} found at $z \sim 1$ a slight correlation between the metallicity gradient and the sSFR  \citep[but see][ for a different view]{Carton2018MNRAS, Ma2017MNRAS}. Here we test if we are able to recover the observed trends using the simplest assumption of a universal metallicity gradient. 

The iron over hydrogen abundance ratio can be estimated for each DLA, for a given impact parameter $b=r$, applying the following formula:
\begin{equation}
\bigg[\frac{\rm Fe}{\rm H} \bigg] (r)= log_{10}(Z_{\rm DLA,0})-log_{10}(Z_{\odot})-\Gamma r
\end{equation}

where 
$$Z_{\rm DLA,0}=Z_{\rm DLA}(r=0)=$$
$$=\frac{\mu_{\rm HI}}{\mu_{\rm Fe}} \frac{{\rm M}_{\rm Fe} (1-11 \, {e}^{-10})}{{\rm M}_{\rm HI}} \frac{(1-<f_{mol}>)}{\int^{10}_0 dy (1-f_{mol} (y))} e^{-(1+ ln(10) \Gamma r_s) y}   $$

and $y=r/r_s$, while $r_s=R_{{\rm CG},d}$ and $f_{mol}$ is the molecular fraction.

\section{Properties of simulated DLAs}
In this section, we compare the properties of the DLAs in our simulated Universe with those estimated from observational data.
For each property derived in this section we have combined the MSI and MSII simulation as explained in Section \ref{Sect:CatalogCreation}.
\subsection{The column density distribution function}
The column density distribution function (CDDF) is defined as the number of absorbers observed per unit redshift path and column density interval:
\begin{equation}
f(N_{\rm HI},X) dX dN_{\rm HI} = n_{abs}(N_{\rm HI},X),
\end{equation}
where the absorbing path $dX$ is defined as
$dX = \frac{H_0}{H(z)} {(1+z)}^2 dz $, in terms of the redshift path $dz$.

The CDDF plays, in absorption line studies, a similarly central role (and provides a similarly `vague' information) as the luminosity function in galaxy evolution studies. 
The analytic model of the CDDF proposed by \cite{schaye01} (devised for over-densities that cannot self-shield from the UV background), together with results from cosmological simulations \citep[e.g.][] {Altay2011}, indicate that systems of a given column density originate from dramatically different over-densities. Nevertheless, higher column densities systems are typically connected to denser gas that, in general and average sense, tends to lie closer to galaxies. It has been argued that the steepest part of the CDDF, made of the densest absorbers, may be particularly sensitive to stellar feedback and stellar evolution \citep{Rosenberg2003ApJ, Bird2014}.

\begin{figure}
\begin{center}
 \includegraphics[width=1.15\columnwidth]{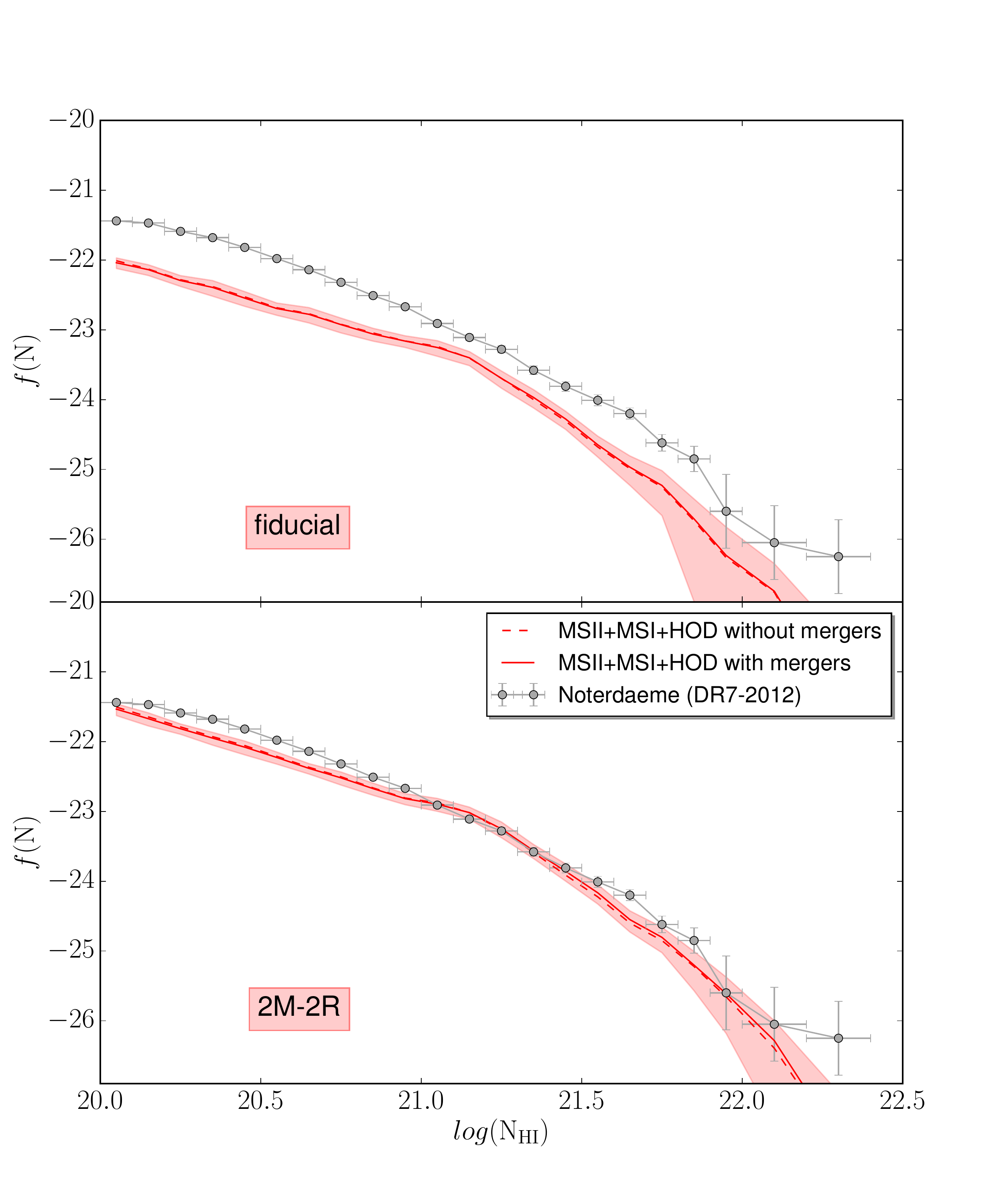}
 \caption{Predicted column density distribution function (CDDF) in the redshift range $2<z<3$. The top panel shows results based on our fiducial model, while the bottom panel shows the results of the model where $R_s=2R_{s,{\rm orig}}$ and $M_{CG}=2M_{CG,{\rm orig}}$.  We estimate the CDDF for each of the 125 realizations described in Sect.~\ref{Sect:CatalogCreation}, converting into a redshift interval ($dz$) the length of each LOS, that is equal to $L_{box}=100 h^{-1} \rm Mpc$ comoving at all redshifts. The solid red line indicates the average of the CDDFs obtained for all realizations considered in the redshift range of interest (2<z<3), while the shaded area
 highlight the $1-\sigma$ scatter of the distribution. 
 }
   \label{fig:CDDF-zrange-fiducial}
\end{center}
\end{figure}

Fig.~\ref{fig:CDDF-zrange-fiducial}, in the upper panel, shows the CDDF derived from our simulated absorbers in the redshift range $2<z<3$, for our fiducial combination of halo mass cuts (see subsection ~\ref{sec:Minimal_HOD_model}) applied to fiducial GAEA model. Model predictions are compared with observational estimates by \citet{Noterdaeme2012}.

%
The figure shows a significant discrepancy between our fiducial model and the observed CDDF, in particular below $\log(N_{\rm HI}) < 21$. This discrepancy motivated us to test the dependence of the CDDF on the physical properties of  simulated galaxies, and in particular the scale radius of the gaseous disk and the cold gas mass. The lower panel of Fig.~\ref{fig:CDDF-zrange-fiducial} shows results obtained multiplying by a factor $2$ both the scale radius and the cold gas mass of all model galaxies (the 2M-2R model introduced above). Results from this ad-hoc modifications are in very good agreement with observations at $20.0 < \log(N_{\rm HI}) < 22.2$, for the redshift range considered. We have verified that the better agreement with observational data is mainly driven by the increase of scale radius, that leads to a larger galaxy cross-section (i.e. a larger probability of intersecting model galaxies). 

\subsection{The cosmic hydrogen density associated with DLAs}

The cosmic hydrogen density associated with DLAs can be computed as:

\begin{equation}
    \Omega_{\rm DLA}=\frac{m_H H_0 \sum_{i} N_i(HI)}{c \rho_c \Delta X}.
\label{eq:Omega-DLA}
\end{equation}
where $\rho_c$ is the critical density at $z=0$, $m_{\rm H}$ is the mass of the hydrogen atom, and the sum is carried out over all systems with $\log {\rm  N (H I )} > 20.3$, across a total absorption path length $\Delta \rm X$.

\begin{figure}
\hspace*{-0.8cm}
 \includegraphics[width=1.2\columnwidth]{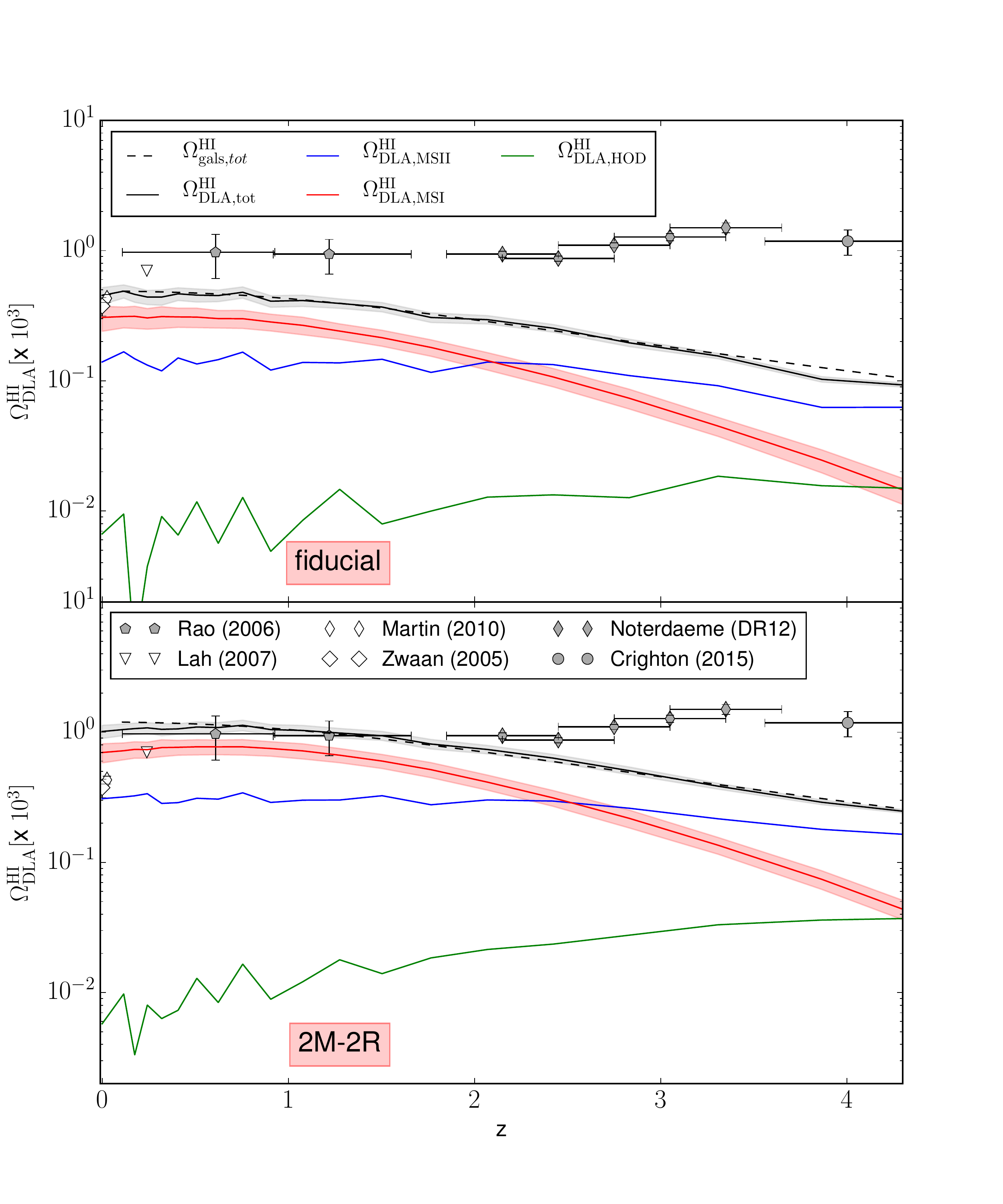}
 \caption{The top (bottom) panel shows the evolution with redshift of $\Omega^{\rm HI}_{\rm DLA}$, in our fiducial (2R-2M) model. We define $\Omega^{\rm HI}_{\rm DLA}= 1.2 \times \Omega_{\rm DLA}$, taking into account the contribution to the comoving HI density of systems with column density lower than the characteristic one of DLAs \citep{Crighton2015}. The solid black line shows the average $\Omega^{\rm HI}_{\rm DLA}$ evolution considering the contribution of all individual systems, while the dashed black line refers to the comoving HI density ($\Omega_{\rm HI}$ ) of all the galaxies in the box.
 The three solid lines in red, blue and green refer to the DLAs in the MSI, MSII and HOD respectively. Symbols with error bars show observational data points, taken from the literature as detailed in the legend, and expressed in the cosmology used by \citet{Crighton2015}.}
  \label{fig:Omega_DLA-evolution}
\end{figure}
Fig.~\ref{fig:Omega_DLA-evolution} compares the redshift evolution of the comoving HI density derived from our simulated DLAs, with the observational estimates from \cite{Crighton2015}. To be consistent with the observations, we have corrected the values provided by Eq.~\ref{eq:Omega-DLA} by a factor $1.2$ ($\Omega^{\rm HI}_{\rm DLA}= 1.2 \times \Omega_{\rm DLA}$), that takes into account the contribution to the comoving HI density of absorbers with column density below $N_{HI} = 20.3$ \citep{Crighton2015}. Using this correction, $\Omega^{\rm HI}_{\rm DLA}$ turns out to agree remarkably well with $\Omega^{\rm HI}_{\rm gals}$ that is derived summing the HI contribution of all model galaxies. This non trivial result indicates that our model predicts the correct shape for the CDDF (both for the fiducial and the $2M-2R$ model).

$\Omega^{\rm HI}_{\rm DLA}$ derived from our simulations ('fiducial' model) is, on average, a factor $\sim 2.5$ below the observational estimates in the redshift range $0 < z < 2$, and it further decreases to about an order of magnitude below the data at $z=4$. As discussed in Sect.~\ref{sec:Properties-simulated-gals}, the difference is in part due to the fact that our model is tuned to reproduce the HIMF measured in the local Universe \citep{Martin2010HIMF}, that gives an estimate of the integrated comoving HI density a factor $\sim 2$ lower than that derived from DLA observations. The 2M-2R assumption alleviates the discrepancy, at least up to redshift $ z \sim 3 $, as shown in the lower panel of Fig.~\ref{fig:Omega_DLA-evolution}. 

At higher redshift, also the predictions from the $2M-2R$ model exhibit a significant decline, while observations measure little evolution of $\Omega^{\rm HI}_{\rm DLA}$ up to $z \sim 5$. 
This can be due to different reasons: one hypothesis is that the uniform redistribution of the missing hydrogen, applied in the $2M-2R$ model, is limited, since it gives too much gas to the more massive haloes, which already reproduce the observations, and too less to the intermediate/ low mass ones. The other possibility is that the contribution of outflows and/or filamentary structure becomes more significant at higher redshift \citep[e.g.][]{VanDeVoort2012DLAs, Fumagalli2011MNRAS}.

\subsection{DLA metallicity}

\subsubsection{Relation between metallicity and $\rm N_{ HI}$}
We compare the metallicity of our simulated DLAs with observations taking advantage of the catalog by \cite{DeCia2018}, that provides also dust-corrected abundance ratios. As explained earlier, we adopt the iron over hydrogen abundance ratio ($\rm [Fe / H]$) as a proxy of the metallicity, and we analyze separately the 125 DLA catalogs built (as described in Section \ref{Sect:CatalogCreation}), in the redshift range $2<z<3$. 

\begin{figure}
\begin{center}
\hspace*{-0.5cm}
 \includegraphics[width=1.15\columnwidth]{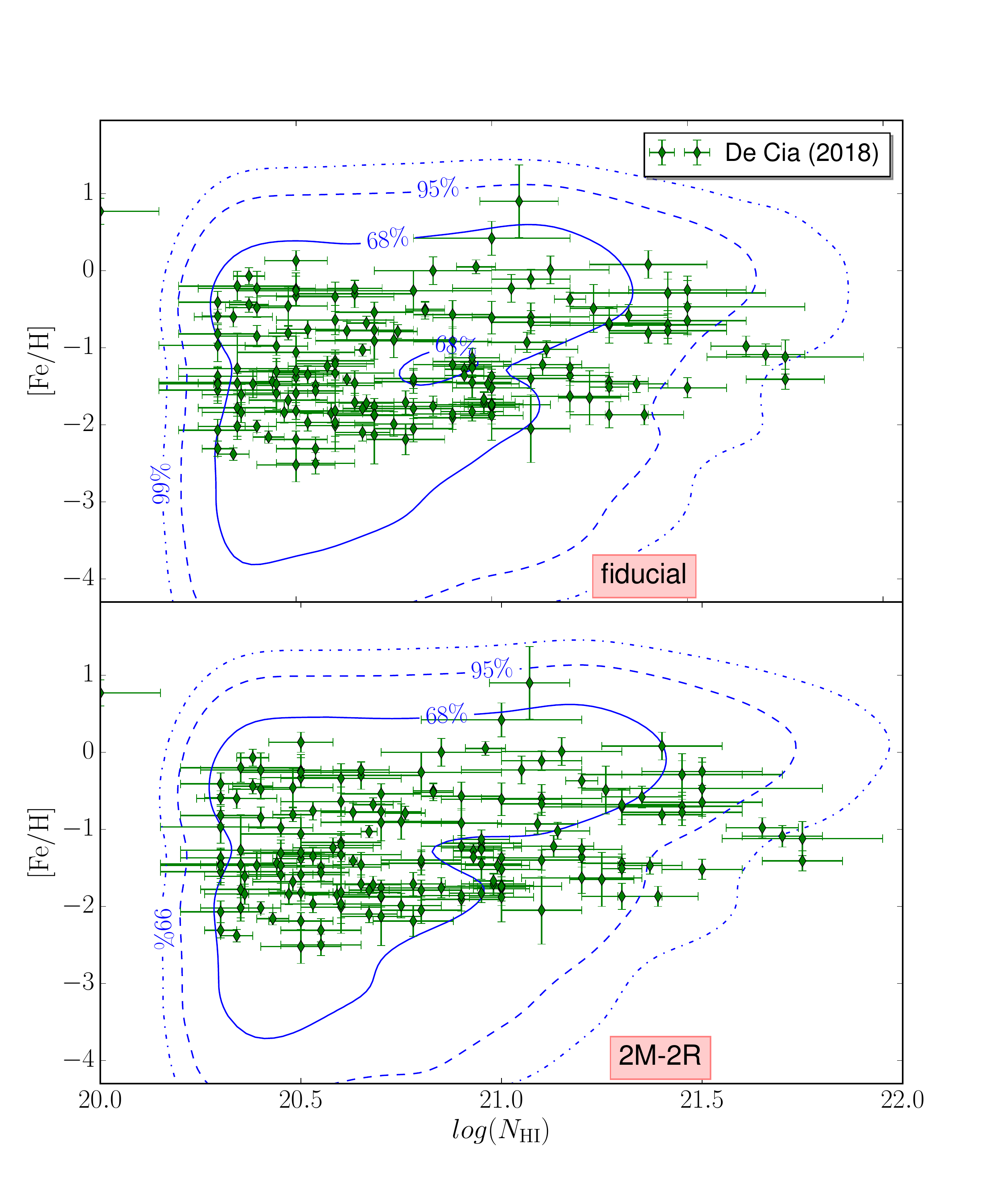}
 \caption{$[\rm \frac{Fe}{H}]$ as a function of $N_{\rm HI}$ in the redshift range $2<z<3$, with no correction for a metallicity gradient. The top and bottom panels show the metallicity distributions based on the fiducial and 2M-2R models, respectively. In both panels, we show the distributions of abundance ratios obtained by stacking the 125 realizations considered.}
 \label{fig:contourplots-abundances}
\end{center}
\end{figure}

\begin{figure}
\begin{center}
\hspace*{-0.8cm}
 \includegraphics[width=1.15\columnwidth]{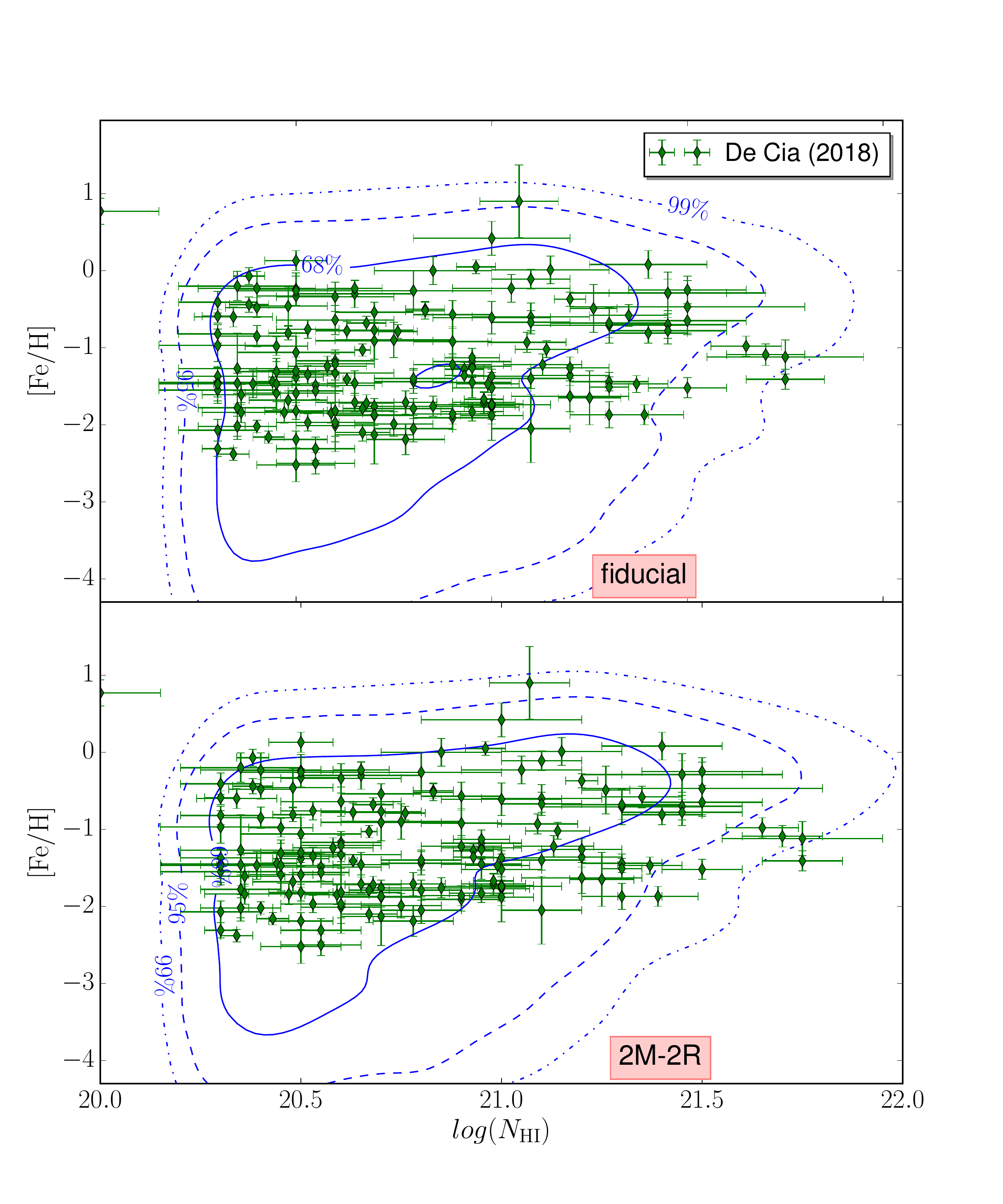}
 \caption{As in Fig.~\ref{fig:contourplots-abundances}, but applying a correction for the metallicity gradient, based on the fitting formula by \citet{Christensen2014MR}.}
 \label{fig:contourplots-abundances-with-gradient}
\end{center}
\end{figure}

Fig.~\ref{fig:contourplots-abundances} and Fig.~\ref{fig:contourplots-abundances-with-gradient} show $\rm [Fe / H]$ as a function of $ \rm N_{ HI}$ in the redshift range $2<z<3$, with lines of different styles contouring the regions enclosing  $68$, $95$ and $99$ per cent of the distribution coming out from the stacking of the 125 DLA  catalogues. Green symbols with error bars show observational measurements. Fig.~\ref{fig:contourplots-abundances} shows results obtained assuming a uniform distribution of the metals in the gaseous disk, while Fig.~\ref{fig:contourplots-abundances-with-gradient} shows the distribution obtained assuming a universal metallicity gradient (see Sect.~\ref{Sect:Metallicity}). 

The simulated abundance ratios appear in somewhat better agreement with the data when we consider a metallicity gradient, in particular at larger metallicity values. Our model, however, predicts a not negligible number of low abundance ratios ($\rm [Fe/H]<-3.$) systems that are not observed (\cite{Prochaska2009}). 

To make the comparison more quantitative, we carry out a two-dimensional Kolmogorov-Smirnov test to quantify the probability that the simulated and observed distributions are extracted from the same sample (i.e. are consistent). 

The estimate of the p-value, namely the probability of obtaining the observed distribution assuming the null hypothesis,
is $1.16 \cdot 10^{-5}$ ($1.32 \cdot 10^{-5}$) for the $2M-2R$ run, with (without) a correction for the metallicity gradient. The corresponding value for the fiducial model is $1.77 \cdot 10^{-7}$ ($5.53 \cdot 10^{-7}$).
Therefore the hypotesis that the observed DLA metallicities come from the same parent population of the simulated DLAs is on average rejected for both  models considered, independently of the metallicity gradient applied.


The discrepancy between observed and simulated data is mainly driven by the presence, in our model, of low-metallicity systems that are absent in the observed DLA samples. In addition, our simulated DLAs corresponding to large column densities tend to have an average metallicity that is larger than the observed one. This `shift' in the average metallicity at higher column densities of simulated systems, with respect to that observed, increases in the 2M-2R model. 

The excess of low-metallicty systems in our model suggests that the treatment of the chemical enrichment of low-mass haloes (see Fig.~\ref{fig:FeH_vs_Mh}) may be inadequate and should be revised. The difference in the distributions at larger metallicities is more difficult to explain. It is worth noting that the High $A_V$ Quasar survey \citep[HAQ][]{Fynbo2013HAQ,Krogager2015HAQ,Zafar2015HAQ} and the extended-HAQ \citep{Krogager2016eHAQ} have shown that the traditional quasar selection used in SDSS is biased against reddened quasars. In addition, the work by \citet{Noterdaeme15} showed that DLAs associated with large column-densities and metallicities are typically found to exhibit a more significant reddening of the background quasar. Therefore, it is plausible that the combined effect of dust and large atomic hydrogen densities cause a dust-bias in DLA observations, preferentially excluding from the observations DLAs hosted in massive, metal-rich and dusty galaxies. In the $2M-2R$ model, the average metallicity is slightly larger than for the fiducial model, due to the reassignment 'a posteriori' of the scale radius and the mass, which penalizes the low-mass galaxies at intermediate/high column density (see Appendix A).

\subsubsection{Cosmic metallicity evolution}

As observed by \cite{Rafelski2012ApJ}, the chemical enrichment of DLAs evolves from about $1$ to $10$ per cent solar from $z\sim 5$ to today. The work by \cite{Rafelski2012ApJ} also revealed a statistically significant decline of the DLA average metallicity 
with increasing redshift, that can be described as $<\Omega_Z> = (−0.26 \pm 0.07) z - (0.59 \pm 0.18)$.
This behaviour  was confirmed at $z<4$  by independent measurements
\citep{Kulkarni2007ApJ,Kulkarni2010NewA}.

\citet{DeCia2018} also found a similar decrease with redshift, but with a different normalization at low redshift, based on abundance ratios corrected for dust depletion.

\begin{figure}
\begin{center}
 \hspace*{-0.5 cm}
 \includegraphics[width=1.1\columnwidth]{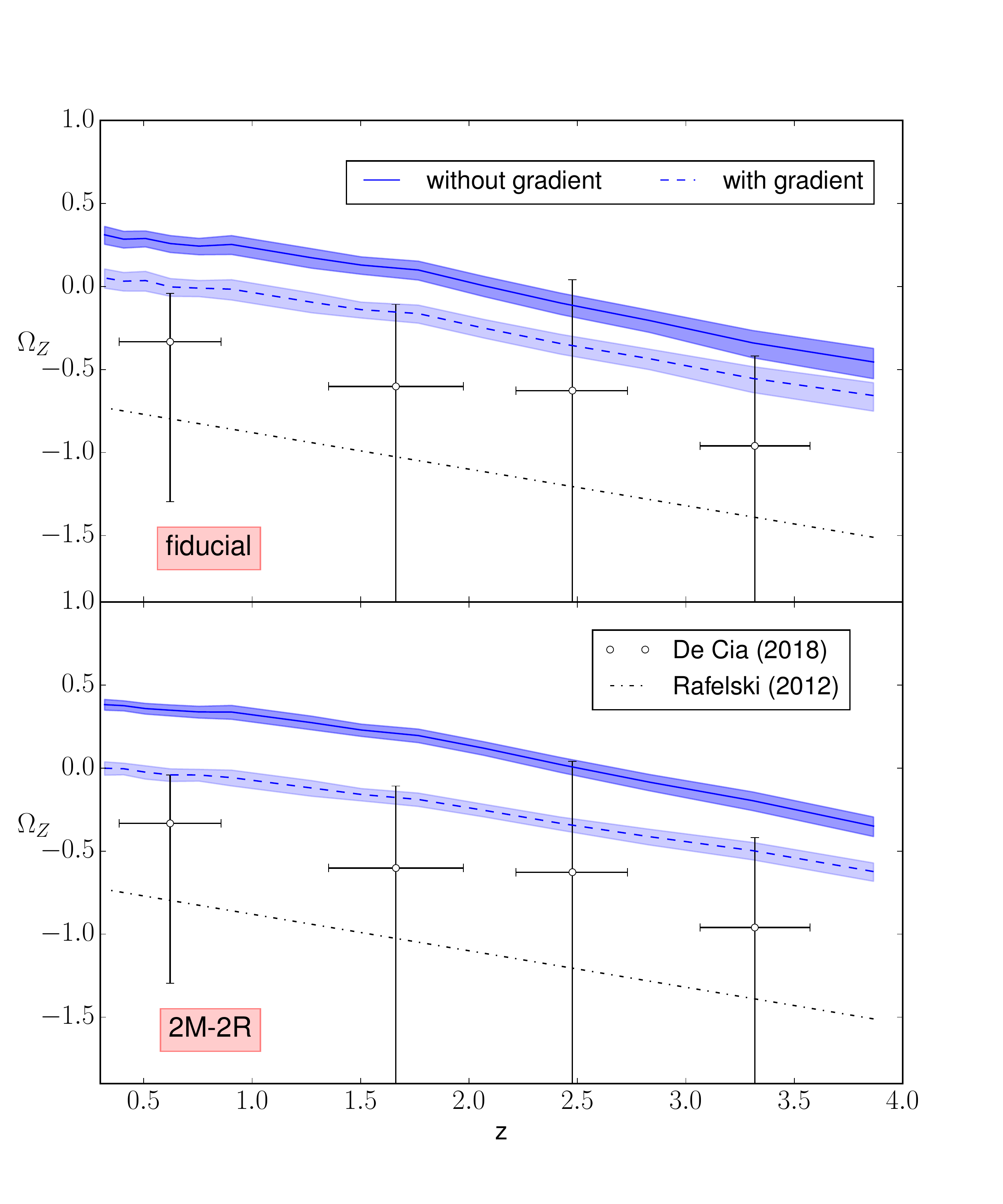}
 \caption{Cosmic metallicity evolution. Solid and dashed blue lines show model predictions without and with a correction for metallicity gradient, respectively. The shaded areas highlight the relative $1- \sigma$ scatter regions. The top panel shows results for our fiducial GAEA run, while the bottom panel corresponds to the 2M-2R model.}
   \label{fig:OmegaZ-evolution}
\end{center}
\end{figure}

We have investigated the evolution of the DLA metallicity by computing the mean cosmic metallicity of simulated DLAs at different redshifts. Following \citet{Rafelski2012ApJ}, this can be defined, at each redshift, as:

\begin{equation}
<\Omega_Z> = log_{10}( {\sum}_i \frac{{10}^{[M/H]_i} N_{HI, i}}{{\sum}_i N_{HI, i}})
\label{eq:Omega-Z}
\end{equation}

where the index $i$ runs over all DLAs in the redshift bin considered, and $[M/H]_i$ is the adopted metal abundance ratio (in our case [Fe/H]). Fig.~\ref{fig:OmegaZ-evolution} shows the mean cosmic metallicity as defined in Eq.~\ref{eq:Omega-Z} for the redshift range $0.3<z<4$, together with the fitting function (black dashed line) found by \cite{Rafelski2012ApJ} and the data from \cite{DeCia2018}. We consider the latter sample as our reference data sample, since our metal abundances do not account for dust depletion.

When we apply a correction for the metallicity radial gradient, the mean evolution of the cosmic metallicity of our simulated DLAs is in agreement with the data by \cite{DeCia2018} within the errors, although model predictions tend to give always higher median values than the median of the data, at all redshifts considered. This is expected because, as discussed above, observations likely miss most of the DLAs at high column densities with large metallicity.

\subsection{DLA host halo masses}

The typical range of halo masses hosting DLAs is still an open question, albeit the low detection rate of DLA counter-parts in optical follow-up observations suggest that DLAs are most likely associated with faint galaxies and therefore reside in small haloes \citep{Fynbo1999MNRAS,Krogager2017}. If we adopt the same argument used in abundance matching studies \citep[e.g.][]{Conroy2006ApJ}, the results by \citet{Fynbo2008ApJ} can be translated into a typical DLA host halo mass of $M_{200} <10^{11}\,M_{\odot}$. This is, however, in tension with more recent observational work based on DLA kinematics and clustering. The distribution of velocity widths measured from low ionization metal lines shows a prominent tail at high velocities, which suggests the existence of a population of large discs hosting DLAs \citep{Bird2015MNRAS}. Moreover, the recent cross-power spectrum analysis by \citet{Font-Ribera2012}, based on the BOSS survey, provides an estimate of the linear bias of the observed DLAs  ($ \rm b_{DLA} = 2.17 \pm 0.20$), suggesting a typical host halo mass $\sim {10}^{12} M_{\odot}$. This analysis has been updated by
\citet{Perez-Rafols2018MNRAS}, who found a linear bias of $\rm b_{DLA} = 2.00 \pm 0.19$, only slightly lower than the clustering amplitude measured for Lyman Break Galaxies \citep[LBGs, see ][]{Cooke2006ApJ}, and no dependence of the bias value on redshift or column density. This bias value implies $M^{\rm DLA}_{\rm host} \geq {10}^{11} M_{\odot}$, that is larger than the one typically predicted by some simulations and semi-analytic models \citep{Pontzen2008,Barnes2014PASP,Padmanbhan2017MNRAS}. The median typical DLA host halo masses, found in the redshift range $2 < z < 3$ from our model, are listed in Table \ref{Table: DLA halo mass per z} and are in agreement with observational results by \citet{Perez-Rafols2018MNRAS}.

Recently, \cite{PerezRafols18} have shown that the bias of DLAs exhibits a dependence on metallicity, in line with preliminary observational results \citep{Neeleman2013ApJ,Christensen2014MR} and the expectation that more metal-rich DLAs are associated with more massive galaxies. In our model, we also see a variation of the average metal content of DLAs hosted in haloes of different masses, and this can be explained as a consequence of the relation between the gas metallicity in galaxies and the host halo mass (see Fig.~\ref{fig:FeH_vs_Mh}).

\begin{table}
\begin{center}
\begin{tabular}{|c|c|c|}

z & $M_{200}$ (fiducial) & $M_{200}$ (2M-2R) $[{10}^{11} M_{\odot}]$ \\
\hline
2.07 &  ${1.50 \pm}^{0.64}_{0.21}$ & $ {2.67 \pm }^{0.56}_{0.50}$  \\
\hline
2.42 & ${1.32 \pm}^{0.14}_{0.23}$ & ${2.00 \pm}^{0.38}_{0.43} $  \\
\hline
3.06 & ${0.71 \pm }^{0.18}_{0.11}$ & ${0.94 \pm}^{0.21}_{0.11}$ \\ 
\hline
\end{tabular}

\caption{Median DLA host halo masses predicted by our fiducial and 2M-2R model.}
\label{Table: DLA halo mass per z}
\end{center}
\end{table}

\section{Discussion}
 
In this work, we have analysed the properties of Damped Lyman$-\alpha$ systems (DLAs) by taking advantage of a semi-analytic model of galaxy formation and evolution \citep[GAEA, presented in][]{Hirschmann2016ADS,Xie2017} coupled to two large cosmological N-body simulations: the Millennium (MSI) and Millennium II (MSII). In order to estimate the possible contribution from haloes that are below the resolution of our simulations, we have used a simple HOD approach by placing, at random positions within the simulated box, a number of haloes with mass distribution consistent with analytic formulations tuned on N-body simulations \citep{Tinker2008}. Our model assumes that all atomic hydrogen is associated with the gaseous disk of galaxies, i.e. there is no contribution from filamentary regions or extraplanar gas. Our simulated DLAs catalogues are then built by throwing a large number (100,000) of random lines of sight along the z-direction of 125 simulated boxes, obtained combining the simulations available and complemented with HOD extrapolation (as described in Section~\ref{Sect:CatalogCreation}). 

Our fiducial model predicts a column density distribution function with the correct shape but offset low with respect to observational measurements by \cite{Noterdaeme2012}. This affects the predicted values of the cosmic hydrogen density in DLAs (${\Omega}^{\rm HI}_{\rm DLA}$), that is a factor $\sim 2.5$ lower than observational estimates at $0<z \leq 2$, and even more at higher redshift. Up to $z<3$ the disagreement with data can be overcome by increasing the radius of the gaseous disk and the gas mass by a factor $\sim 2$ (our $2M-2R$ model). As for the DLA metallicity distribution in the redshift range $2 <z < 3$, our model predicts an excess of low metallicity DLA systems,  
while the average cosmic DLA metallicity ($\Omega_{\rm Z}$), weighted over $\rm N_{HI}$, follows the same redshift evolution as observational measurements but it is slightly higher than observed values. The predicted $\Omega_{\rm Z}$ becomes compatible with observations, within the uncertainties, once we account for a modest radial metallicity gradient.

Below, we discuss our results in relation with independent recent studies, and point out possible developments/improvements of the adopted physical model that can bring model results in better agreement with observational measurements. 

\subsection{Comparison with the literature}

In the last twenty years, a number of theoretical studies, either using a semi-analytic approach \citep{Lagos2011MNRAS, Lagos2014MNRAS, Berry2014MNRAS, Kim2015-HI-mass-function-Photoionization-low-mass-end} or hydro-dynamical simulations \citep{Nagamine2007ApJ, Pontzen2008, Tescari2009MNRAS, Altay2011, Cen2012ApJ}, have focused on the evolution of the atomic hydrogen content of the Universe. 

In the framework of this paper, it is particularly interesting to discuss our results in relation to the analysis by \citet{Berry2014MNRAS,Berry2016MNRAS}, also based on semi-analytic models and focused on the predicted properties of DLAs. In their work, \cite{Berry2014MNRAS} use variations (see their Table 1) of the semi-analytic model published in \citet[][see also \citealt{Popping2014MNRAS,Somerville2015MNRAS}]{Somerville2008}, including different prescriptions for the partition of cold gas in atomic and molecular hydrogen, and alternative assumptions for the sizes of gaseous disks.  
Our model and the one used by \cite{Berry2014MNRAS} differ significantly for the numerical implementation
and for the prescriptions adopted for modelling various physical processes.  
\citet{Popping2014MNRAS} and \citet{Xie2017} show that both models are able to reproduce the evolution of disc sizes (both stellar and gaseous) up to $z\sim 2$, for galaxies more massive than $10^9 M_{\odot}$.
Fig.~2 of \cite{Berry2014MNRAS} shows that none of the model variants they considered reproduces well the local HI mass function, while our fiducial model is tuned to reproduce this observational constraint. 

Both our fiducial run and the reference disc model used in \citet{Berry2014MNRAS} under-predict the column density distribution function of DLAs. \citet{Berry2014MNRAS} find a better agreement by increasing the cold gas specific angular momentum with respect to what assumed in their reference model. This leads to larger gaseous disks, but also to a significantly worse agreement with the HI galaxy mass function in the local Universe (see their Fig.2). This is consistent with our findings that a model where we arbitrarily multiply by a factor two both the scale radii and HI masses of model galaxies better reproduces the observed column density distribution function. Ours is an `ad-hoc' solution, and it remains to be demonstrated that plausible modifications of the modelled physical processes can lead to such solution without (significantly) affecting the agreement shown between model predictions and observational data in the local Universe. We will come back to this issue in the next section, in the framework of possible developments of the GAEA model. 

It should be noted that also our 2M-2R model, that reproduces the observed column density distribution of DLAs for $z<3$, predicts a decline of ${\Omega}^{\rm HI}_{\rm DLA}$ at higher redshift. This is in disagreement with observational measurements and consistent with what found by \cite{Berry2014MNRAS,Berry2016MNRAS}. This decline is driven by an under-estimation of the column density distribution function for $log (N_{\rm HI})<21$. The behaviour is not shared by hydro-dynamical simulations that typically not underestimate the CDDF for $log (N_{\rm HI})<21$ and find no evolution or even a moderate increase of ${\Omega}^{\rm HI}_{\rm DLA}$ \citep{Cen2012ApJ,VanDeVoort2011MNRAS,Altay2011}, in better agreement with observational measurements. 

The different behavior at high redshift, predicted by semi-analytic models and hydro-dynamical simulations, could be at least in part explained by an increasing contribution to the DLAs cross-section of filamentary structures and outflows/inflows at higher redshift \citep{VanDeVoort2011MNRAS,Fumagalli2011MNRAS,Cen2012ApJ}. 
In addition, simulations predict that at $z \sim 3$ the halos that contribute most to the CDDF for $log (N_{\rm HI})<21$ are the ones in the mass range ${10}^{9}< M_{200} < {10}^{10} M_{\odot}$ \citep{Tescari2009MNRAS,Rahmati2013} while in our model the major contribution comes from halos in the mass range ${10}^{10}< M_{200} < {10}^{12} M_{\odot}$. Since more massive halos are less numerous at higher redshift, the difference in the typical DLA host halo mass at $log (N_{\rm HI})<21$  could partially explain the decline of the ${\Omega}^{\rm HI}_{\rm DLA}$ in our model.
Another concern is related to the possible contribution of haloes that are below the resolution of our simulations. In order to understand to what extent low-mass haloes contribute to the HI comoving density, we have estimated the contribution of haloes with mass $\rm {10}^{8}M_{\odot} <M_{200}< {10}^{9.2} M_{\odot}$ resorting to a simple HOD model (see Sec.\ref{sec:Properties-simulated-gals} for details). Our results indicate that these low-mass haloes represent a negligible contribution to the column density distribution in the redshift range of interest.
The average covering fraction of HI in different halos is influenced also by the interplay between the UV background and the gas density in the galactic disks.
At the column density typical of DLA systems, the gas is self-shielded by the ionizing photons of the UV background, then mostly neutral. Our semi-analytical model does not include a specific treatment for the self-shielding but this effect is taken into account implicitly through the adoption of the BR prescription for the cold gas partitioning \citep{Blitz-Rosolowsky2006}. Albeit the \cite{Blitz-Rosolowsky2006} prescription is based on observations of local galaxies, we are confident in applying this prescription to all redshifts, since we have demonstrated that it provides very similar results to alternative parametrizations based e.g. on hydro-dynamical simulations that account explicitly for self-shielding \citep{Xie2017}.\\
As discussed in the previous sections, both our fiducial and $2M-2R$ models predict an excess of low-metallicity DLAs that are not present in observational samples, also in the dust-corrected DLA abundance ratio catalog by \cite{DeCia2018}. Results based on the model by \citet{Somerville2015MNRAS} appear in better agreement with the observed metallicity distribution of DLAs \citep[see Fig.10 in][]{Berry2014MNRAS}. This difference is likely due to the different treatment adopted for the metal enrichment. In particular, the Somerville model assumes an instantaneous recycling approximation and sets a metallicity floor for the hot gas in low mass haloes (the haloes with $M_{vir} \leq {10}^9 M_{\odot}$ are set to have a hot gas metallicity equal to ${10}^{-3} Z_{solar}$ ). Our model instead does not assume pre-enrichment of gas in low-mass haloes and includes a detailed chemical enrichment scheme that accounts for the non instantaneous recycling of gas and metals \citep{DeLucia2014MNRAS}. In addition, as discussed above, we assume that the $95 \%$ of newly synthesized metals is directly injected into the hot gas phase in low mass haloes \citep{Hirschmann2016ADS}, which contributes to delay the chemical enrichment of low-mass systems.
Fig. 16 of \cite{Somerville2015MNRAS} shows that their model predicts almost no evolution with redshift of the mass-metallicity relation while our model predicts an increasing normalization at lower redshift.  The different redshift evolution together with the different slope (less steep for low-mass galaxies in the case of the Somerville model) could lead to gas metallicities, for galaxies in the mass range ${10}^7 < M_{\star} < {10}^8$, that are larger
in the Somerville et al. model
than in ours. This could also contribute to the different predictions obtained for the metallicity distribution of DLAs, in particular at low metallicities.   


For the typical DLA host halo mass, predictions from our model (both for the fiducial and the $2M-2R$ run) are similar to those by the Somerville model. 
For a mean redshift $z=2.3$ (taking all DLAs with $1.97\leq z \leq 2.6$) we find a median DLA host halo mass equal to $M^{DLA}_{host} = 1.55 \times {10}^{11} M_{\odot}$ for the fiducial model and $M^{DLA}_{host} = 2.28 \times {10}^{11} M_{\odot}$ for the $2M-2R$ model, in agreement with observational estimates by \cite{Perez-Rafols2018MNRAS}.

\subsection{Model developments}

Our results suggest possible avenues to improve the
agreement between the predictions of the GAEA model and observational data of DLAs:
(i) increasing the HI content of model galaxies; (ii) increasing the sizes of gaseous disks; and (iii) modifying the treatment for the metal enrichment of low-mass haloes. In this section, we discuss plausible implementations that can bring the model in this direction. In future work, we intend to explore these suggestions in more detail.  

The HI content of model galaxies depends on the assumed prescription for cold gas partitioning. In our model, the molecular to atomic hydrogen ratio is slightly larger than what observed in the local Universe by xGASS and xCOLDGASS \citep{Catinella2018a,Saintonge2017xCOLDGASS}. This is shown and discussed in a forthcoming paper \citep[][]{Xie2020arXiv200312757}. Naively, one could think that a lower molecular fraction can be obtained by simply increasing the star formation efficiency: stars are formed from molecular gas and larger star formation rates should lead to consume more molecular hydrogen. The situation is, however, complicated by the strong self-regulation between star formation and stellar-feedback that makes model results not very sensitive to the star formation law adopted \citep[][and references therein]{Xie2017}. In addition, simple modifications of model parameters would generally require a retuning of the model to restore the agreement with the main observables used as constraints (in our case the HI mass function).  

Another possible reason for the too low HI masses of galaxies in intermediate mass haloes is the prescription adopted for reionization. Our model assumes an `early' reionization (with starting redshift $z_0=15 $ and completed by $z_{r} \sim 11$) that is inconsistent with recent Planck results \citep{Planck2018results}. The reionization feedback is implemented through a `filtering mass' whose evolution is described by the analytic fitting function introduced in \citet{Kravtsov2004ApJ} \citep[based on the simulation results by][]{Gnedin2000ApJ}. 
Adopting a time-line for reionization more in agreement with recent results, 
we expect a filtering mass lower by an order of magnitude with respect to the one assumed in our model for $z>5$.
Besides, \citet{Okamoto2008} showed that the parametrization of the filtering mass presented in \cite{Gnedin2000ApJ}, based on low-resolution simulations, might over-estimate by up to one order of magnitude (at $z=0$) the characteristic mass where photo-ionization feedback becomes effective in reducing the baryon fraction (see their Fig.~6), independently of the assumed reionization history. 

The size of the HI galactic disks in our model is determined by the evolution of the specific angular momentum of the cold gas.
\cite{Xie2020arXiv200312757} have significantly updated
the treatment of the angular momentum, leading to both larger gaseous disks and larger HI masses, in the direction of the 2M-2R model considered in previous sections.
It is worth noting that also the SFR sizes predicted by our fiducial model tend to be smaller than observational estimates \citep{Xie2017}, and also this disagreement is relieved with the larger disks obtained with the updated angular momentum scheme \citep[as shown in,][]{Xie2020arXiv200312757}.

Finally, the excess of low-metallicity DLAs, with respect to observational measurements, can possibly be solved by modifying the fraction of metals that are injected directly into the hot gas component in low-mass haloes. This will likely affect also the cooling times (the cooling function is very sensitive to the metallicity of the hot gas), leading to a lower gas accretion rate onto galaxies, but could be compensated by the above-described modifications
concerning the reionization scenario.
  
\section{Summary}

In this work we have investigated the properties of Damped Lyman$-\alpha$  systems (DLAs) taking advantage of the state-of-the-art semi-analytical model GAEA \citep[our fiducial model is the BR run, described in][]{Xie2017}. We have used model outputs obtained by considering two large cosmological simulations: the Millennium \citep[MSI,][]{Springel_MS2005} and the Millennium II simulation \citep[MSII,][]{BoylanMSII2009MNRAS}, with higher resolution but smaller box. We consider also the contribution of DM haloes below the resolution of the simulation adopting a simple HOD approach and populating the box with isolated DM haloes in the mass range ${10}^8< M_{200}< {10}^9 M_{\odot}$.

From the comparison of the GAEA model predictions with DLA observations in the redshift range $2<z<3$, we find that the fiducial GAEA model reproduces the overall shape of the column density distribution function (CDDF), but predicts a CDDF and an ${\Omega}^{\rm HI}_{\rm DLA}$ that are offset systematically below the observational measurements. The agreement with observations is significantly improved, at least up to $z \sim 3$, increasing "a posteriori" both the HI masses and the gaseous disk radii of model galaxies by a factor 2 (we have referred to this as the 2M-2R model in the text).
At higher redshift ($z>3$) our predicted $\Omega^{\rm HI}_{\rm DLA}$ decreases in both the model versions considered, in disagreement with observational measurements and differently from what happens in hydrodynamical simulations.

Our analysis of the relative contribution to the DLA comoving HI density of simulated DM haloes, in bins of virial mass, highlights that DM haloes with $M_{200}<10^9 M_{\odot}$ (under the resolution of the adopted N-body simulations) do not give a significative contribution to $\Omega_{\rm HI}$ up to $z\sim 4$, in the framework of our study.

Our model predicts a population of DLAs with very low abundance ratios ($[Fe/H]<-2.5$), not in agreement with the observed metallicity floor \citep{Storrie-Wolfe2000ApJ,Prochaska2009,Rafelski2012ApJ}, suggesting to explore the possibility of a modification of our prescription for the metal ejection in low mass haloes.
At the same time, the simulated DLA with high column density have an average metallicity larger than the observed one, leading to a consequent relatively high normalization of the values of the cosmic DLA metallicity ($\Omega_Z$). This discrepancy is mitigated when a correction accounting for the metallicity gradient in galaxies, based on the fitting formula by \citet{Christensen2014MR}, is applied.

Our model predicts a median DLA halo host mass of $\sim {10}^{11} M_{\odot}$, in agreement with the results of the work by \cite{Perez-Rafols2018MNRAS} on the DLA Lyman-$\alpha$ cross-correlation analysis.

The predicted DLA impact parameters have a distribution which is also in agreement with the estimates derived by \cite{Krogager2017}, in particular in the case of the model with larger galactic discs ($2M-2R$ model).

The picture emerging from the present analysis, which includes a detailed comparison with the similar work by \cite{Berry2014MNRAS,Berry2016MNRAS}, 
suggests possible improvements of the physical prescriptions of our model, in particular
the HI content of galaxies, the sizes of gaseous disks
and the metal enrichment of low-mass haloes. 

\section*{Data availability}
The data underlying this article will be shared on reasonable request to the corresponding author, Serafina Di Gioia.

\section*{Acknowledgements}

SDG thanks Valentina D'Odorico for helpful answers on DLA observations, Giorgio Calderone and Emiliano Munari for helping in making the analysis code faster, and Anna Zoldan and Marta Spinelli for insightful discussion on HI in galaxies and semi-analytical models. SC acknowledges financial support from PRIN-MIUR 201278X4FL: {\it Evolution of cosmic baryons: astrophysical effects and growth of cosmic structures}. 




\bibliography{biblioSera} 



\begin{figure}
\begin{center}
\begin{minipage}[c]{\textwidth}
 \includegraphics[width=18cm]{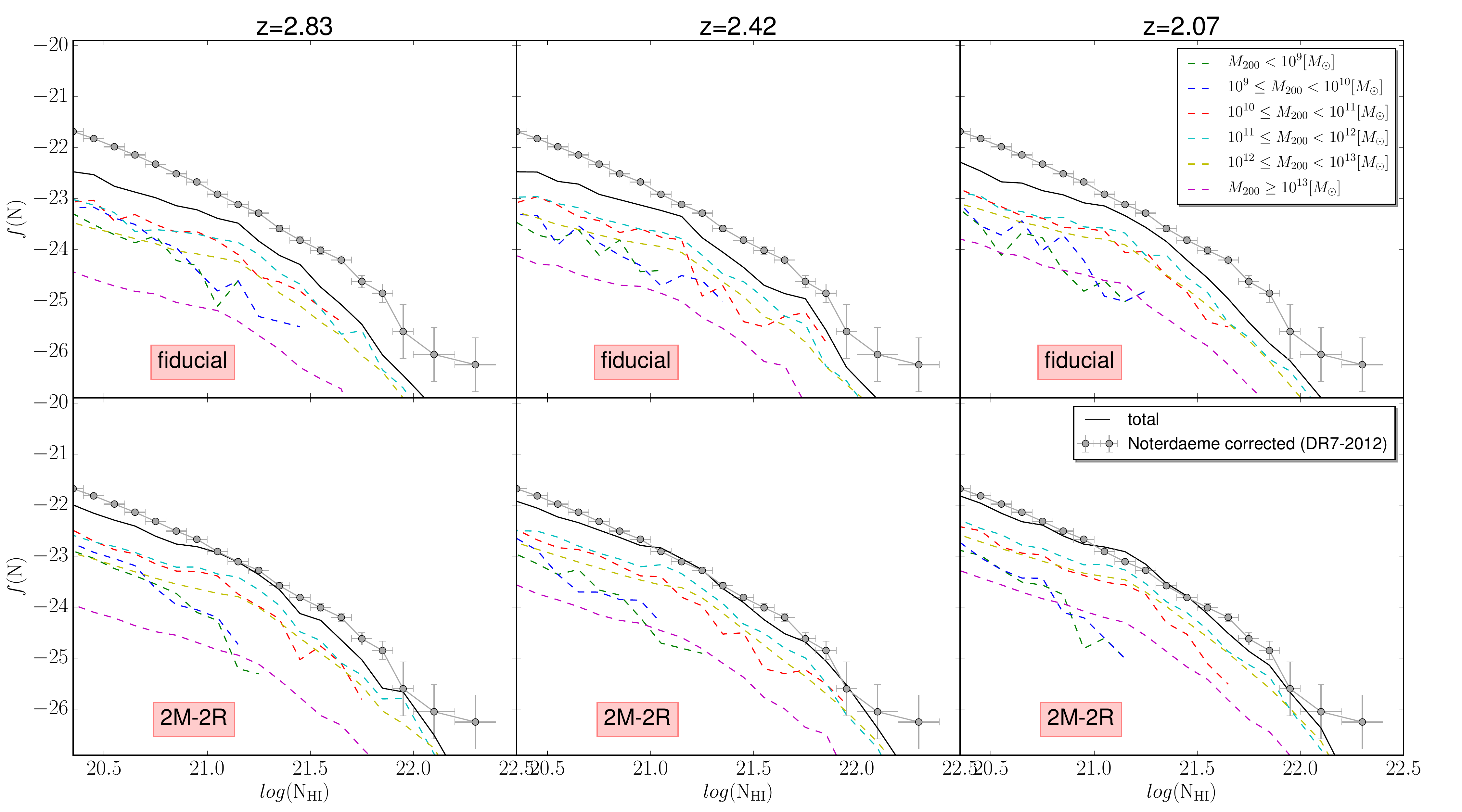}
 \caption{Evolution with redshift of the predicted CDDF and its dependence on the DLA host halo masses. The top panels show results from our fiducial model, while the bottom panels show the corresponding results from the 2M-2R model. The black solid lines show the total CDDF, while the dashed colored lines show the average contribution to the CDDF of dark matter haloes in different virial mass bins, as indicated in the legend.}
\label{fig:CDDF_different_z_binned_Mvir}
\end{minipage}
\end{center}
\end{figure}
\begin{figure}
\begin{center}
 \includegraphics[width=1.2\columnwidth]{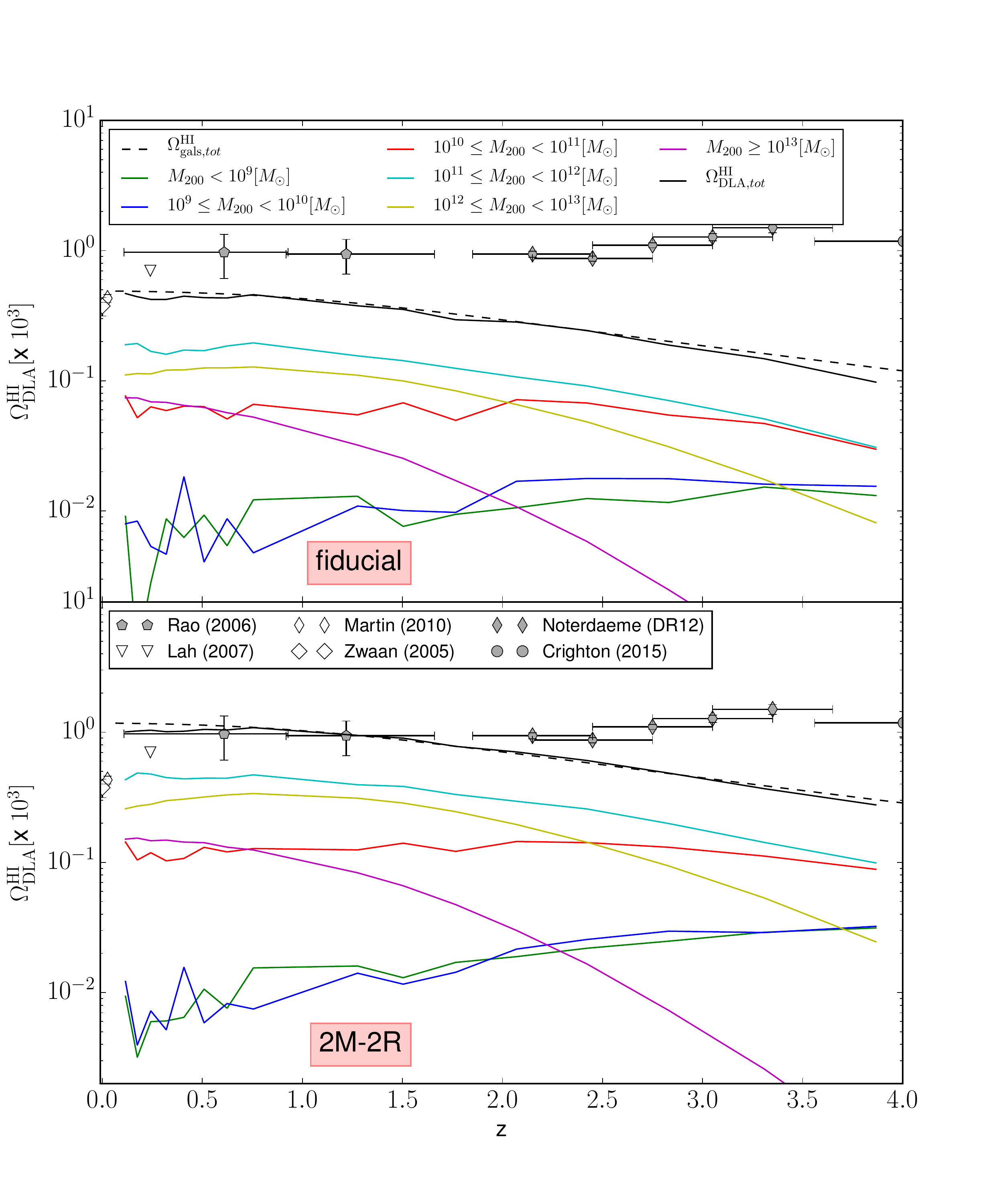}
 \caption{Evolution with redshift of the predicted comoving HI density in DLAs ($\Omega^{\rm DLA}_{\rm HI}$) and its dependence on the DLA host halo masses. The top panel shows results from our fiducial model, while the bottom panel shows the corresponding results from the 2M-2R model. The black solid lines show the total $\Omega^{\rm DLA}_{\rm HI}$, while the dashed colored lines show the average contribution to $\Omega^{\rm DLA}_{\rm HI}$ of dark matter haloes in different virial mass bins, as indicated in the legend.}
\label{fig:OmegaDLA_binned_Mvir}
\end{center}
\end{figure}

\appendix
\clearpage
\begin{center}
\begin{minipage}[c]{\textwidth} 

\section{Evolution of the DLA statistics with redshift and in different mass bins}

Observations indicate a negligible evolution of the CDDF as a function of redshift \citep{Noterdaeme2012}, while other theoretical studies based on hydrodynamical simulations \citep{Rahmati2013} found little evolution of the low column density end, with the slope becoming steeper at higher redshift. In contrast with observations, our model predicts a moderate evolution of the CDDF, in particular of the low column density end, that flattens at lower z.
To understand the origin of this evolution it can be useful to investigate how different DLA host halo masses are distributed in different column density bins.

Fig. \ref{fig:CDDF_different_z_binned_Mvir} shows the predicted CDDF at three different redshifts ($z=2.83, \, 2.42, \, 2.07$), with the dashed lines highlighting the contribution of haloes of different mass, and the bottom (top) panel showing the results of our $2M-2R$ (fiducial) model.

Haloes in the mass bin ${10}^{11} M_{\odot} \leq M_{200} < {10}^{12} M_{\odot}$ represent the major contribution to the CDDF at all column densities - a contribution that decreases at higher redshift, as expected in a hierarchical scenario. The second major contribution come from haloes in the mass bin ${10}^{10} M_{\odot} \leq M_{200} < {10}^{11} M_{\odot}$ for intermediate/low DLA column densities and from haloes in the mass bin ${10}^{12} M_{\odot} \leq M_{200} < {10}^{13} M_{\odot}$ for high DLA column densities.
Haloes with $M_{200} < {10}^{10} M_{\odot}$ start to contribute significantly for $log({\rm N_{HI}}) < 20.7 $ around $z \sim 2.8$ and moving to higher redshift their relative contribute to all column densities increases.

In the $2M-2R$ model the contribution to the CDDF of the haloes with $M_{200}>{10}^{11} M_{\odot}$ increases at all column densities while the contribution of low mass haloes ($M_{200}<{10}^{10} M_{\odot}$) increases only at low column densities (for $N_{\rm HI} < 21$) and that of the intermediate mass (${10}^{10} M_{\odot}<M_{200}<{10}^{11} M_{\odot}$) haloes increases for intermediate column densities (up to $N_{\rm HI}=1.4$), with respect to the fiducial model. This could be explained remembering that the HI surface density scales linearly with the mass and as the inverse of the square radius of the galactic disk. 
Moreover, since the cross-section increases quadratically with the galaxy scale radius, a larger number of disks are intersected by our l.o.s. in the 2M-2R model, in particular at larger halo masses. Therefore, the predicted CDDF from this model is in better agreement with observational measurements. 

In Fig. \ref{fig:OmegaDLA_binned_Mvir} we show, for the redshift range $0 < z < 4$ the contributions to the predicted $\Omega_{\rm DLA}^{\rm HI}$ of haloes in different mass bins, with logarithmic bin size $\Delta \log(M_{200}/M_{\odot}) = 1. $
The haloes which contribute more at all redshifts are those in the mass bin ${10}^{11} M_{\odot} \leq M_{200} < {10}^{12} M_{\odot}$. The second largest contribution is provided by the mass bin ${10}^{12} M_{\odot} \leq M_{200} < {10}^{13} M_{\odot}$ up to $z=2.5$ and by the mass bin ${10}^{10} M_{\odot} \leq M_{200} < {10}^{11} M_{\odot}$ for $z>2.5$.
It is worth noting that the contributions of the two lowest mass bins are very similar and both represent less than $10\%$ of the total $\Omega_{\rm DLA}^{\rm HI}$ in the entire redshift range considered. 

\end{minipage}
\end{center} 

\clearpage

\clearpage

\section{Influence of different gas vertical density profiles on the estimated DLA properties}
The distribution of HI gas detected through the 21 cm line is fairly flat and uniform \citep{Leroy2008AJ}, with a scalelength much larger than stellar disk one.
The work by \cite{Narayan2002A&A} suggests that the vertical structure of the gaseous disk is sensitive to the gravity of all galactic components, i.e. stars, dark matter and gas.
Under the assumption of an isothermal distribution, one expects that the gaseous/stellar vertical density profile is described by the function  ${sech}^2$, as shown theoretically by \cite{Spitzer1942ApJ} and confirmed by some observations \citep[][]{VanderKruit1982A&A...110...61V}.
However, more recent observational studies have found that the observed vertical distribution for gas and stars in galaxies is steeper than the one predicted by an isothermal distribution, and it
is well-approximated by an exponential or a $sech$ function, especially close to the galactic
mid-plane \citep{Barteldrees1994A&AS..103..475B,Rice1996AJ}  

We have considered the effect on the DLA column density distribution function (CDDF) of assuming a different vertical density profile for the model galaxies.

We assume 4 different density profiles for the gas in the galactic disc:
the 'classic' double-exponential 
$$\rho_{\rm CG}(r,z)=\rho_0 ~ e^{-{r}/{R_s}} ~e^{-{z}/{z_0}} $$
and three additional profiles, described by the formula presented in \cite{VanDerKruit2011}:

$$ \rho_{\rm CG}(r,z)=\rho_0 ~ e^{-{r}/{R_s}} sech^{\frac{2}{n}}(\frac{ n \, {\rm z}}{2 z_0})$$
with $n=1,2$ and $ 4$ respectively.

\begin{figure}
\begin{center}
 \includegraphics[width=1.1\columnwidth]{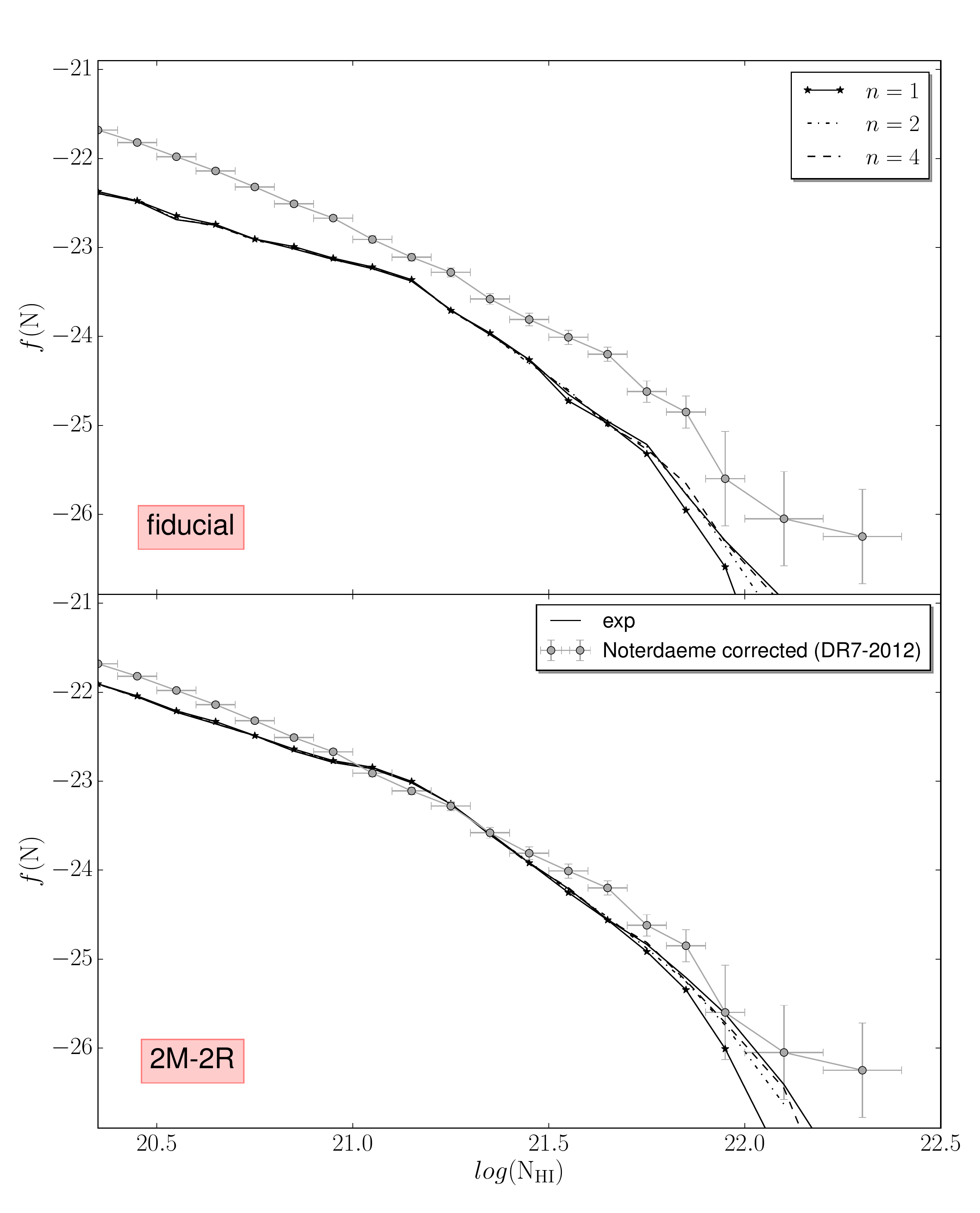}
 \caption{Average CDDF in the redshift range $2<z<3$, compared to the data (grey dots) by \citet{Noterdaeme2012}. The top (bottom) panel shows the results from the fiducial ($2M-2R$) model. The black lines describe the total CDDF while the other lines show the average contribution to the CDDF assuming different density profiles, as described in the legend.}
 \label{fig:CDDF_average_z_vprofiles}
\end{center}
\end{figure}

In Fig \ref{fig:CDDF_average_z_vprofiles} the predicted average CDDF in the redshift range 2<z<3 is shown. The solid line refers to the exponential vertical profile, while the star-dashed, dot-dashed and dashed line refer to the function presented in \cite{VanDerKruit2011}, respectively with $n=1, \, 2, \,4$.

Fig \ref{fig:CDDF_average_z_vprofiles} highlights that the 4 different density profiles lead to differences in the CDDF only in the high column-density regime. And for large $n$-values the class of functions presented by \cite{VanDerKruit2011} give very similar results to the exponential density profile, for all the column densities considered.

\bsp	
\label{lastpage}

\end{document}